\documentclass{aastex701}

\usepackage{amsmath}

\newcommand{\sigmaru}{\sigma_{\text{ARU}}}
\defcitealias{lhaasomrk4212026}{The LHAASO Collaboration, 2026}
\defcitealias{hessEBL2013}{The H.E.S.S. Collaboration, 2013}
\defcitealias{hessM87}{The H.E.S.S. Collaboration, 2024}

\begin{document}

\title{Constraining the Photon Intensity of Extragalactic Background Light with the HAWC Observatory for the Blazar Mrk 421}



\author{R.~Alfaro}
\email{ruben@fisica.unam.mx}
\affiliation{Instituto de F\'{i}sica, Universidad Nacional Autónoma de México, Ciudad de Mexico, Mexico}%

\author{C.~Alvarez}
\email{crabpulsar@hotmail.com}
\affiliation{
    Universidad Autónoma de Chiapas, Tuxtla Gutiérrez, Chiapas, México
}%

\author{A.~Andrés}
\email{alexis.andres1210@gmail.com}
\affiliation{Instituto de Astronom\'{i}a, Universidad Nacional Autónoma de México, Ciudad de Mexico, Mexico}%

\author{E.~Anita-Rangel}
\email{earangel@astro.unam.mx}
\affiliation{%
    Instituto de Astronom\'{i}a, Universidad Nacional Autónoma de México, Ciudad de Mexico, Mexico
}%

\author[orcid=0000-0002-0595-9267]{M.~Araya}
\email{miguel.araya@ucr.ac.cr}
\affiliation{%
    Universidad de Costa Rica, San José 2060, Costa Rica
}%

\author{J.C.~Arteaga-Velázquez}
\email{juan.arteaga@umich.mx}
\affiliation{Universidad Michoacana de San Nicolás de Hidalgo, Morelia, Mexico}%

\author[orcid=0000-0002-4020-4142]{D.~Avila Rojas}
\email{daniel_avila5@ciencias.unam.mx}
\affiliation{%
    Instituto de Astronom\'{i}a, Universidad Nacional Autónoma de México, Ciudad de Mexico, Mexico
}%

\author{R.~Babu}
\email{baburish@msu.edu}
\affiliation{Department of Physics and Astronomy, Michigan State University, East Lansing, MI, USA}%

\author[orcid=0000-0002-3886-3739]{P.~Bangale}
\email{priyadarshini.bangale@temple.edu}
\affiliation{%
    Temple University, Department of Physics, 1925 N. 12th Street, Philadelphia, PA 19122, USA
}%

\author{E.~Belmont-Moreno}
\email{belmont@fisica.unam.mx}
\affiliation{Instituto de F\'{i}sica, Universidad Nacional Autónoma de México, Ciudad de Mexico, Mexico }%

\author{A.~Bernal}
\email{abel@astro.unam.mx}
\affiliation{%
    Instituto de Astronom\'{i}a, Universidad Nacional Autónoma de México, Ciudad de Mexico, Mexico
}%

\author[orcid=0000-0003-2158-2292]{T.~Capistrán}
\email{tcapistranc@gmail.com}
\affiliation{%
    Instituto Nacional de Astrof\'{i}sica, Óptica y Electrónica, Puebla, Mexico
}%

\author[orcid=0000-0002-8553-3302]{A.~Carramiñana}
\email{alberto@inaoep.mx}
\affiliation{%
    Instituto Nacional de Astrof\'{i}sica, Óptica y Electrónica, Puebla, Mexico
}%

\author{F.~Carreón}
\email{mfcarreon@astro.unam.mx}
\affiliation{%
    Instituto de Astronom\'{i}a, Universidad Nacional Autónoma de México, Ciudad de Mexico, Mexico
}%

\author{A.L.~Colmenero-Cesar}
\email{1615807e@umich.mx}
\affiliation{Universidad Michoacana de San Nicolás de Hidalgo, Morelia, Mexico}%

\author[orcid=0000-0002-7607-9582]{U.~Cotti}
\email{umberto.cotti@umich.mx}
\affiliation{%
    Universidad Michoacana de San Nicolás de Hidalgo, Morelia, Mexico
}%

\author{J.~Cotzomi}
\email{jcotzomi@yahoo.com.mx}
\affiliation{Facultad de Ciencias F\'{i}sico Matemáticas, Benemérita Universidad Autónoma de Puebla, Puebla, Mexico }%

\author{S.~Coutiño de León}
\email{sara.cdl989@gmail.com}
\affiliation{Dept. of Physics and Wisconsin IceCube Particle Astrophysics Center, University of Wisconsin{\textemdash}Madison, Madison, WI, USA}%

\author[orcid=0000-0002-7574-1298]{N.~Di Lalla}
\email{niccolo.dilalla@stanford.edu}
\affiliation{%
    Department of Physics, Stanford University: Stanford, CA 94305--4060, USA
}%

\author[orcid=0000-0001-8487-0836]{R.~Diaz Hernandez}
\email{dihera77@gmail.com}
\affiliation{%
    Instituto Nacional de Astrof\'{i}sica, Óptica y Electrónica, Puebla, Mexico
}%

\author[orcid=0000-0001-8451-7450]{B.L.~Dingus}
\email{dingus@lanl.gov}
\affiliation{%
    Los Alamos National Laboratory, Los Alamos, NM, USA
}%

\author{M.A.~DuVernois}
\email{duvernois@icecube.wisc.edu}
\affiliation{%
    Dept. of Physics and Wisconsin IceCube Particle Astrophysics Center, University of Wisconsin{\textemdash}Madison, Madison, WI, USA
}%

\author{T.~Ergin}
\email{ergin.tulun@gmail.com}
\affiliation{Department of Physics and Astronomy, Michigan State University, East Lansing, MI, USA }%

\author[orcid=0000-0001-7074-1726]{C.~Espinoza}
\email{m.catalina@fisica.unam.mx}
\affiliation{%
    Instituto de F\'{i}sica, Universidad Nacional Autónoma de México, Ciudad de Mexico, Mexico
}%

\author[orcid=0000-0002-0173-6453]{N.~Fraija}
\email{nifraija@astro.unam.mx}
\affiliation{%
    Instituto de Astronom\'{i}a, Universidad Nacional Autónoma de México, Ciudad de Mexico, Mexico
}%

\author{S.~Fraija}
\email{sarafraija@hotmail.com}
\affiliation{%
    Instituto de Astronom\'{i}a, Universidad Nacional Autónoma de México, Ciudad de Mexico, Mexico
}%

\author[orcid=0000-0002-4188-5584]{J.A.~García-González}
\email{anteus79@gmail.com}
\affiliation{%
    Tecnologico de Monterrey, Escuela de Ingenier\'{i}a y Ciencias, Ave. Eugenio Garza Sada 2501, Monterrey, N.L., Mexico, 64849
}%

\author[orcid=0000-0003-1122-4168]{F.~Garfias}
\email{fergar@astro.unam.mx}
\affiliation{%
    Instituto de Astronom\'{i}a, Universidad Nacional Autónoma de México, Ciudad de Mexico, Mexico
}%

\author{J.A.~González}
\email{jose.gonzalez.c@umich.mx}
\affiliation{%
    Universidad Michoacana de San Nicolás de Hidalgo, Morelia, Mexico
}%

\author{N.~Ghosh}
\email{nghosh1@mtu.edu}
\affiliation{Department of Physics, Michigan Technological University, Houghton, MI, USA }%

\author{A.~Gonzalez Muñoz}
\email{adiv.gonzalez@itoaxaca.edu.mx}
\affiliation{Instituto de F\'{i}sica, Universidad Nacional Autónoma de México, Ciudad de Mexico, Mexico }%

\author{M.M.~González}
\email{magda@astro.unam.mx}
\affiliation{Instituto de Astronom\'{i}a, Universidad Nacional Autónoma de México, Ciudad de Mexico, Mexico }%

\author{J.A.~Goodman}
\email{goodman@umd.edu}
\affiliation{Department of Physics, University of Maryland, College Park, MD, USA }%

\author{S.~Groetsch}
\email{sjgroets@mtu.edu}
\affiliation{%
    Department of Physics, Michigan Technological University, Houghton, MI, USA
}%

\author{J.~Gyeong}
\email{kyoungjh1011@naver.com}
\affiliation{%
    Department of Physics, Sungkyunkwan University, Suwon 16419, South Korea
}%

\author[orcid=0000-0002-2565-8365]{S.~Hernández-Cadena}
\email{shkdna@sjtu.edu.cn}
\affiliation{%
    Tsung-Dao Lee Institute \& School of Physics and Astronomy, Shanghai Jiao Tong University, 800 Dongchuan Rd, Shanghai, SH 200240, China
}%

\author[orcid=0000-0001-5169-723X]{I.~Herzog}
\email{herzogia@msu.edu}
\affiliation{%
    Department of Physics and Astronomy, Michigan State University, East Lansing, MI, USA
}%

\author[orcid=0000-0002-5527-7141]{F.~Hueyotl-Zahuantitla}
\email{filihz@gmail.com}
\affiliation{%
    Universidad Autónoma de Chiapas, Tuxtla Gutiérrez, Chiapas, México
}%

\author{D.~Huang}
\email{dezhih@mtu.edu}
\affiliation{University of Delaware, Department of Physics and Astronomy, Newark, DE, USA}

\author[orcid=0000-0001-5811-5167]{A.~Iriarte}
\email{airiarte@astro.unam.mx}
\affiliation{%
    Instituto de Astronom\'{i}a, Universidad Nacional Autónoma de México, Ciudad de Mexico, Mexico
}%

\author{S.~Kaufmann}
\email{skaufmann13@googlemail.com}
\affiliation{%
    Universidad Politecnica de Pachuca, Pachuca, Hgo, Mexico
}%

\author[orcid=0000-0003-4785-0101]{D.~Kieda}
\email{dave.kieda@utah.edu}
\affiliation{%
    Department of Physics and Astronomy, University of Utah, Salt Lake City, UT, USA
}%

\author[orcid=0000-0001-5516-4975]{H.~León Vargas}
\email{hleonvar@fisica.unam.mx}
\affiliation{%
    Instituto de F\'{i}sica, Universidad Nacional Autónoma de México, Ciudad de Mexico, Mexico
}%

\author[orcid=0000-0001-8825-3624]{A.L.~Longinotti}
\email{alonginotti@astro.unam.mx}
\affiliation{%
    Instituto de Astronom\'{i}a, Universidad Nacional Autónoma de México, Ciudad de Mexico, Mexico
}%

\author{G.~Luis-Raya}
\email{gilura6969@hotmail.com}
\affiliation{Universidad Politecnica de Pachuca, Pachuca, Hgo, Mexico }%

\author{K.~Malone}
\email{kmalone@lanl.gov}
\affiliation{Los Alamos National Laboratory, Los Alamos, NM, USA }%

\author{O.~Martinez}
\email{omartin@fcfm.buap.mx}
\affiliation{Facultad de Ciencias F\'{i}sico Matemáticas, Benemérita Universidad Autónoma de Puebla, Puebla, Mexico }%

\author[orcid=0000-0002-2824-3544]{J.~Martínez-Castro}
\email{macj@cic.ipn.mx}
\affiliation{%
    Centro de Investigaci\'on en Computaci\'on, Instituto Polit\'ecnico Nacional, M\'exico City, M\'exico.
}%

\author{H.~Martínez-Huerta}
\email{humberto.martinezhuerta@udem.edu}
\affiliation{Departamento de Física y Matemáticas, Universidad de Monterrey}%

\author{P.~Miranda-Romagnoli}
\email{pa.miranda.r@gmail.com}
\affiliation{Universidad Autónoma del Estado de Hidalgo, Pachuca, Mexico }%

\author{P.E.~Mirón-Enriquez}
\email{pelimi92@gmail.com}
\affiliation{%
    Instituto de Astronom\'{i}a, Universidad Nacional Autónoma de México, Ciudad de Mexico, Mexico
}%

\author{E.~Moreno}
\email{emoreno@fcfm.buap.mx}
\affiliation{Facultad de Ciencias F\'{i}sico Matemáticas, Benemérita Universidad Autónoma de Puebla, Puebla, Mexico }%

\author{M.~Mostafá}
\email{miguel@psu.edu}
\affiliation{Temple University, Department of Physics, 1925 N. 12th Street, Philadelphia, PA 19122, USA}%

\author{M.~Najafi}
\email{mnajafi@mtu.edu}
\affiliation{Department of Physics, Michigan Technological University, Houghton, MI, USA }%

\author{L.~Nellen}
\email{lukas@nucleares.unam.mx}
\affiliation{Instituto de Ciencias Nucleares, Universidad Nacional Autónoma de Mexico, Ciudad de Mexico, Mexico }%

\author{M.U.~Nisa}
\email{nisamehr@msu.edu}
\affiliation{Department of Physics and Astronomy, Michigan State University, East Lansing, MI, USA }%

\author[orcid=0000-0001-7099-108X]{R.~Noriega-Papaqui}
\email{ropapaqui@gmail.com}
\affiliation{%
    Universidad Autónoma del Estado de Hidalgo, Pachuca, Mexico
}%

\author{E.~Ponce}
\email{eponce@fcfm.buap.mx}
\affiliation{%
    Facultad de Ciencias F\'{i}sico Matemáticas, Benemérita Universidad Autónoma de Puebla, Puebla, Mexico
}%

\author[orcid=0000-0002-8774-8147]{Y.~Pérez Araujo}
\email{yuniorpy@gmail.com}
\affiliation{%
    Instituto de F\'{i}sica, Universidad Nacional Autónoma de México, Ciudad de Mexico, Mexico
}%

\author[orcid=0000-0001-5998-4938]{E.G.~Pérez-Pérez}
\email{egperezp@yahoo.com.mx}
\affiliation{%
    Universidad Politecnica de Pachuca, Pachuca, Hgo, Mexico
}%

\author[orcid=0000-0002-8940-5316]{A.~Pratts}
\email{yoba_m_t_a@ciencias.unam.mx}
\affiliation{%
    Instituto de F\'{i}sica, Universidad Nacional Autónoma de México, Ciudad de Mexico, Mexico
}%

\author{C.D.~Rho}
\email{no397@naver.com}
\affiliation{Department of Physics, Sungkyunkwan University, Suwon 16419, South Korea}%

\author[orcid=0000-0003-1327-0838]{D.~Rosa-González}
\email{danrosa@inaoep.mx}
\affiliation{%
    Instituto Nacional de Astrof\'{i}sica, Óptica y Electrónica, Puebla, Mexico
}%

\author[orcid=0000-0002-4204-5026]{M.~Roth}
\email{mattroth@lanl.gov}
\affiliation{%
    Los Alamos National Laboratory, Los Alamos, NM, USA
}%

\author[orcid=0000-0001-6079-2722]{A.~Sandoval}
\email{asandoval@fisica.unam.mx}
\affiliation{%
    Instituto de F\'{i}sica, Universidad Nacional Autónoma de México, Ciudad de Mexico, Mexico
}%

\author{M.~Shin}
\email{	minjishin23@gmail.com}
\affiliation{%
    Department of Physics, Sungkyunkwan University, Suwon 16419, South Korea
}%

\author{A.J.~Smith}
\email{asmith8@umd.edu}
\affiliation{Department of Physics, University of Maryland, College Park, MD, USA }%

\author{Y.~Son}
\email{youngwan.son@cern.ch}
\affiliation{%
    University of Seoul, Seoul, Rep. of Korea
}%

\author[orcid=0000-0002-1492-0380]{R.W.~Springer}
\email{wayne.springer@utah.edu}
\affiliation{%
    Department of Physics and Astronomy, University of Utah, Salt Lake City, UT, USA
}%

\author{O.~Tibolla}
\email{omar.tibolla@gmail.com}
\affiliation{Universidad Politecnica de Pachuca, Pachuca, Hgo, Mexico }%

\author[orcid=0000-0002-1689-3945]{I.~Torres}
\email{ibrahim.torres23@gmail.com}
\affiliation{%
    Instituto Nacional de Astrof\'{i}sica, Óptica y Electrónica, Puebla, Mexico
}%

\author[orcid=0000-0002-7102-3352]{R.~Torres-Escobedo}
\email{torresramiro350@sjtu.edu.cn}
\affiliation{%
    Tsung-Dao Lee Institute \& School of Physics and Astronomy, Shanghai Jiao Tong University, 800 Dongchuan Rd, Shanghai, SH 200240, China
}%

\author[orcid=0000-0003-0715-7513]{E.~Varela}
\email{enrique.varela@correo.buap.mx}
\affiliation{%
    Facultad de Ciencias F\'{i}sico Matemáticas, Benemérita Universidad Autónoma de Puebla, Puebla, Mexico
}%

\author[orcid=0000-0001-6876-2800]{L.~Villaseñor}
\email{lvillasen@gmail.com}
\affiliation{%
    Facultad de Ciencias F\'{i}sico Matemáticas, Benemérita Universidad Autónoma de Puebla, Puebla, Mexico
}%

\author{X.~Wang}
\email{xiaojiewang@mst.edu}
\affiliation{%
    Department of Physics, Missouri University of Science and Technology, Rolla, MO, US
}%

\author[orcid=0000-0003-2141-3413]{I.J.~Watson}
\email{ian.james.watson@cern.ch}
\affiliation{%
    University of Seoul, Seoul, Rep. of Korea
}%

\author{H.~Wu}
\email{hwu298@wisc.edu}
\affiliation{Dept. of Physics and Wisconsin IceCube Particle Astrophysics Center, University of Wisconsin{\textemdash}Madison, Madison, WI, USA}%

\author[orcid=0009-0006-3520-3993]{S.~Yu}
\email{sjy5345@psu.edu}
\affiliation{%
    Department of Physics, Pennsylvania State University, University Park, PA, USA
}%

\author{X.~Zhang}
\email{xiyingzhangxyz@gmail.com}
\affiliation{%
    Institute of Nuclear Physics Polish Academy of Sciences, PL-31342 IFJ-PAN, Krakow, Poland
}%

\author[orcid=0000-0003-0513-3841]{H.~Zhou}
\email{hao_zhou@sjtu.edu.cn}
\affiliation{%
    Tsung-Dao Lee Institute \& School of Physics and Astronomy, Shanghai Jiao Tong University, 800 Dongchuan Rd, Shanghai, SH 200240, China
}%

\author[0000-0002-8528-9573]{C.~de León}
\email{cederik.de.leon@umich.mx}
\affiliation{%
    Universidad Michoacana de San Nicolás de Hidalgo, Morelia, Mexico
}%

\collaboration{all}{The HAWC Collaboration}\noaffiliation

\correspondingauthor{Sergio Hernández Cadena}
\email{shkdna@sjtu.edu.cn}
\correspondingauthor{Ramiro Torres Escobedo}
\email{torresramiro350@sjtu.edu.cn}
\correspondingauthor{Hao Zhou}
\email{hao\_zhou@sjtu.edu.cn}

\begin{abstract}
    The blazar Mrk 421 exhibits rapid variability over a wide range of timescales. 
    Spectral differences have been observed during the different emission states of Mrk 421. During the high emission states, tests to constraint the Hubble constant and the photon intensity of Extragalactic Background Light (EBL) can be performed.
    The HAWC observatory provides an exceptionally long term monitoring of the source at TeV energies. We selected periods of high emission state and low emission state in data with total observation time of 2460 transits from the HAWC observatory using the All-sky Root around in an Unbiased way methodology. 
    We report on evidence of a cutoff in the spectrum of Mrk 421 during high emission states. An Exponential Cutoff Power Law is preferred over a Simple Power Law at a $3.8\,\sigma$ level. In the Exponential Cutoff Power Law, the cutoff is found at $13\pm3~\text{TeV}$. Using this result, we provide upper limits on the specific intensity of EBL photons. Moreover, the value of the energy cutoff found in our analysis  is different from the  cutoff expected by the interaction of gamma-rays with EBL photons. This result indicates that the cutoff is intrinsic to the source.
\end{abstract}


\section{\label{sec:intro}Introduction}

Blazars are a subclass of Active Galactic Nuclei (AGN) whose beamed emission aligns near the line of sight of the observer. Markarian (Mrk) 421 is often referred to as the standard candle for blazars, as it has been the subject of numerous observation campaigns across the entire electromagnetic spectrum for the past few decades. Such observations, in conjunction with data from many other AGNs, have helped us to lay the foundations of our current knowledge about the acceleration and radiation processes that explain the observed emission from these extragalactic objects. Mrk 421 is located at a redshift of 0.0308 and R.A. =166.11, DEC = 38.120 (J2000) \citep{Fey_2004}, and the observational data show that Mrk 421 has a complex behavior in terms of temporal evolution and Spectral Energy Distribution (SED).

Mrk 421 shows rapid variations in its spectrum during periods of increased activity. Such variations have timescales of $\thicksim15~\text{min}$ \citep{magicMrk4212010,veritasMrk4212011} with flux enhancements up to one order of magnitude brighter than the average emission lasting over 5 or more days. Additionally, High Emission States (HES) and Low Emission States (LES) exhibit spectral differences between them. Typically, a hardening of the spectrum is observed when using HES and LES observations \citep{magicMrk4212007,magicMrk4212010,veritasMrk4212011,hawcMrk4212025}, as expected from leptonic models \citep{magicMrk4212010,tanami}. For the period of increased activity in 2006, the MAGIC collaboration showed that for a Log-Parabola (LP) spectral model, the position of LP's peak shifted to higher energies as the flux increased, also consistent with leptonic models \citep{magicMrk4212010}. An analysis assuming an Exponential Cutoff Power Law (ECPL) model also indicates that the cutoff energy varies between HES and LES periods \citep{magicMrk4212010}. Recently, the LHAASO collaboration has also reported the existence of a cutoff in the long-term spectrum of Mrk 421 after integrating data with an approximated duration of 4 years  (from 2021 to 2024, see \citetalias{lhaasomrk4212026}).

The SED of Mrk 421 shows two humps with peaks in keV and GeV energy bands that are well explained by one-zone Synchrotron Self Compton (SSC) models, as indicated by multiwavelength analyses of quasi-simultaneous data \citep{lhaaso_mrk421,veritasMrk4212011}. These observations also demonstrate that the TeV gamma-ray  emission region traces back to the base of the jet with a size of twice the Schwarzschild radius of the central black hole\footnote{The Schwarzschild radius of Mrk 421 is estimated to have a size of $2.9\times10^{13}~\text{cm}$, \citep{Srofmrk421}} \citep{veritasMrk4212011}. However, the origin of the emission at lower frequencies, i.e., optical and radio, remains a matter of discussion. Temporal correlation between different bands indicate that optical and radio emissions originate from different regions \citep{VarFACT,taylor2026swift20yearsmultiwavelength}, or that other radiation mechanisms are in play \citep{veritasMrk4212011}. Although optical observations exhibit episodes of emission with periodic behavior at the scale of $\thicksim10$ days \citep{mrk421_20152016_low_emission,veritasMrk4212011}, the null temporal correlation between optical and X-ray/TeV gamma-ray data could also imply that some level of periodicity is observed only at optical due to non-ballistical helical motion in the jet \citep{Rieger_geometrical}. On the contrary, the temporal correlation between X-ray and TeV gamma-ray emissions seems to hold generally \citep{veritasMrk4212011,hawcMrk4212025,magicMrk4212007} indicating that the same particle population is responsible for both emissions \citep{VarFACT}; however, reported X-ray and TeV gamma-ray orphan flares require further explanation \citep{veritasMrk4212011}.

The gamma rays emitted by AGNs are attenuated during their journey to Earth through interactions with other background radiation fields via $e^-/e^+$ pair production, which has a non-zero probability of photon absorption. This absorption creates an observable cutoff, more abrupt than an exponential form, in the SED of AGNs and other extragalactic sources. For gamma rays at GeV and TeV energies, the radiation field responsible for this attenuation is the Extragalactic Background Light (EBL). The EBL radiation field, with wavelengths in the range from 0.1 to 1000 $\mu\text{m}$, is the sum of the cosmic optical and infrared radiation from the emission of stars and galaxies since the reionization epoch \citep{Biteau_2015}. However, the exact EBL photon intensity $\nu I_\nu$ is not well known, as direct measurements suffer from large uncertainties due to galactic foregrounds \citep{Biteau_2015,franceschini_IFEBL}.

Another interesting feature of Mrk 421's SED is the evidence of a cutoff in the short-term and long-term averaged\footnote{The definition of short-term and long-term spectra depends on the experiment doing the observation. In this paper, we refer to short-term periods as observations with integration times less than or equal to 1 month, while for long-term periods, we refer to integration times over scales greater than 1 yr.} intrinsic spectra \citep{magicMrk4212007,magicMrk4212010,magicMrk4212012,hawcMrk4212022}. The MAGIC Collaboration tested LP and ECPL models to fit observations with a total exposure of $\thicksim11.51~\text{hours}$. They found that the ECPL model has a probability $\geq$99\% of being the intrinsic spectrum over a Simple Power Law (SPL) scenario \citep{magicMrk4212010}. VERITAS confirms the existence of a cutoff in the integrated spectrum over 47 hours  \citep{veritasMrk4212011}. The intrinsic spectrum obtained from the long-term monitoring by HAWC using $\thicksim1038~\text{days}$ also indicates the presence of a cutoff around $5~\text{TeV}$ \citep{hawcMrk4212022}. The origin of the cutoff itself is not yet totally understood; however, it is believed that it is intrinsic to Mrk 421 and does not originate from interactions of TeV gamma rays with the EBL. The presence of the intrinsic cutoff  may be attributed to an extra attenuation effect (e.g., from external radiation fields near to the gamma-ray emission zone) and/or an energy budget limit of the accelerator responsible for the TeV gamma rays. Nevertheless, the SED of Mrk 421 exhibits a cutoff with a temporal evolution observationally established.

 Observations of AGNs at TeV scales can provide complementary constraints on the magnitude of EBL attenuation (\citealp{Biteau_2015,franceschini_IFEBL,Greaux_2024}, \citetalias{hessEBL2013,hessM87}, \citealp{hawcebl2022}). Such estimates require an accurate depiction of the intrinsic spectrum, which is not always possible when using the long-term average spectrum. It is dominated by the low-emission state of the observed AGNs and is close to the sensitivity threshold of current experiments. Better estimates of the intrinsic spectrum are possible with data collected during HES, which also improves constraints on other physical processes, attenuation effects, or exotic physics.

In this work, we take advantage of the long term coverage provided by the HAWC observatory after 10 years to select periods of increased activity for the emission of Mrk 421 to constrain the specific intensity of EBL photons, $\nu I_\nu$ (\citealp{franceschini_IFEBL}, \citetalias{hessM87}). In particular, we focus on the periods of HES for Mrk 421 to derive an estimate of its SED and $\nu I_\nu$ in the band of 0.52 to 29.40 $\mu \rm{m}$\footnote{The initial and ending points of this band are estimated using the relation $\lambda_{\text{EBL}} = 1.24\times(E_\gamma/(1~\text{TeV}))~\mu\text{m}$ \citep{franceschini_IFEBL} for gamma rays with energies in the range from $\thicksim400~\text{GeV}$ to $24~\text{TeV}$.}. During HES periods, emission from Mrk 421 reaches energies around tens of TeV and provides important insights not only on the SED but also on the EBL, allowing us to differentiate between numerical models. This represents a major improvement with respect to the previous study by HAWC \citep{hawcebl2022}, where  the averaged long-term SED of Mrk 421 and Mrk 501 is utilized for a similar analysis. A major source of uncertainty in understanding the intrinsic spectrum stems from the limited information about acceleration and radiation mechanisms in AGNs. Here, we adopt a phenomenological approach and leave a more in-depth study of the SED for a future publication.

The paper is organized in the following way. In Section~\ref{sec:hawc} we briefly introduce the HAWC observatory. The criteria to select periods of high emission activity is discussed in Section~\ref{sec:aru}. The details of data selection are discussed in Section~\ref{sec:data}. The model is outlined in Section~\ref{sec:anna} and we present our results in Section~\ref{sec:results}. Finally, we conclude by summarizing and discussing the physical interpretation of our results in Section~\ref{sec:discussion}.

\section{The HAWC Observatory} \label{sec:hawc}

The High Altitude Water Cherenkov (HAWC) observatory is a wide-field surveying facility located on Volcán Sierra Negra (Volcán Tliltépetl), Puebla, Mexico, at an altitude of 4,100 m. The primary detector comprises an array of 300 water Cherenkov detectors (WCDs). Each WCD unit measures 7.3 m in diameter and 5 m in height and contains 200,000 liters of ultra-purified water and is equipped with four photomultiplier tubes (PMTs), with a central $10^{\prime\prime}$ PMT surrounded by three peripheral 8'' PMTs. The main array provides an effective area of 22,000 m$^2$ that allows HAWC to monitor 2/3 of the northern sky every day \citep{nimpaper}. With the introduction of Pass 5 \citep{hawcPass5}, the sensitivity of HAWC is enhanced and provides an ideal instrument for the monitoring of transient objects.

We use the data comprising a total of 2460 transits to select data grouped by fraction of detector triggered during an event (fHit) \citep{hawcPass5} and quarter-decade energy bins in log-scale for estimated energy from 0.316\,--\,316 TeV. In this analysis, we use the Neural Network (NN)  estimator, introduced in \citep{abeysekaraMeasurementCrabNebula2019}, to estimate the energy of the events. The selection using fHit is to determine those periods with increased and low activity using the method ARU to perform high activity state searches (see Section~\ref{sec:aru}). See section \ref{sec:data} for more details about the data selection.

\section{ARU Algorithm}
\label{sec:aru}

The All-sky Root around in an Unbiased way (ARU) algorithm \citep{aruicrc} adapts the \textit{Fermi}-LAT All-sky Variability (FAVA) methodology used to construct the 1FAV and 2FAV catalogs \citep{Ackermann_2013,abdollahiSecondCatalogFlaring2017} to the case of wide field-of-view (FOV) gamma-ray observatories.
A brief description of ARU is as follows. 

\begin{figure}[ht]
    \centering
    \includegraphics[width=0.9\linewidth]{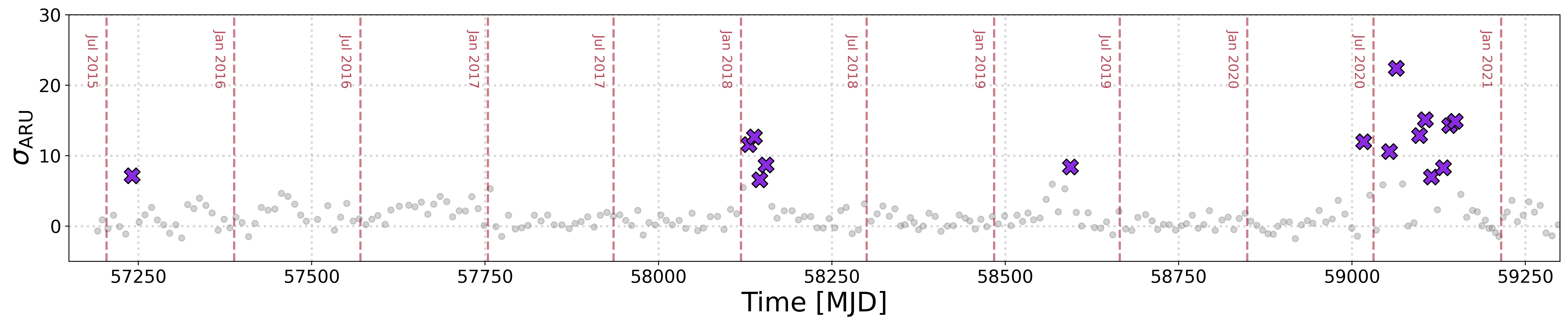}
    \caption{Weekly ARU significance curve for Mrk 421. Significance $\sigmaru$ of deviation of the observed counts from the scaled long-term emission for weekly periods. Purple crosses show all the weekly periods where $\sigmaru$ is greater than 6 and are likely related to periods of increased activity of Mrk 421. Gray  circles indicate periods with $\sigmaru$ smaller than 6.}
    \label{fig:aru_mrk421}
\end{figure}

Assuming that the long-term gamma-ray emission observed at TeV scales is dominated by the constant, steady emission for each source or specific region of the sky,
we can estimate the expected counts at smaller time windows and compare with observations for a given short-term period. The significance $\sigmaru$ provides a way to estimate the level of deviation of the observed counts from the expected emission for a short-term period, and it is obtained from the test statistic $TS_{\text{aru}}$:

\begin{equation}\label{eq:aru_ts}
    TS_{\text{aru}} = -2\ln\left(\frac{\text{Pois}(n^{\text{ST}}_{\text{obs}}|\kappa_{\text{scal}})}{\text{Pois}(n^{\text{ST}}_{\text{obs}}|\kappa_{\text{obs}})}\right),
\end{equation}
with $n^{\text{ST}}_{\text{obs}}$, the observed counts for a short-term period. $\kappa_{\text{scal}}$ is the expected counts scaled from the long-term emission using a ratio of the exposure between short-term and long-term maps, assuming only the steady component from all sources (including all possible contributions without modeling of diffuse gamma-ray emission). $\kappa_{\text{obs}}$ is the observed counts for a short-term that also considers the contribution from flaring and transient sources, $\kappa_{\text{obs}} = \kappa_{\text{scal}} + \delta_\gamma$, with $\delta_\gamma$ containing all the possible transient emissions. In reality, modeling $\delta_\gamma$ is difficult, and we assume that, after fitting the data, $\kappa_{\text{obs}}$ is equal to $n^{\text{ST}}_{\text{obs}}$. $\text{Pois}(n|\kappa)$ is the probability of getting $n$ from a Poisson distribution with mean equal to $\kappa$. Our $TS$ is well defined and described by a $\chi^2$ distribution with the number of degrees of freedom equal to the number of sky maps used in the analysis. The significance $\sigmaru$ is computed assuming a Gaussian transformation. Note that under this transformation,  $TS=0$ corresponds to $\sigmaru = -\infty$, and $\sigmaru = 0$ represents the average deviation from expected short-term emission. A significance of $\sigmaru = -\infty$ is improbable, but all negative values of $\sigmaru$ indicate short-term periods with counts closest to the expected emission.

Our threshold for an increase in emission within a region of interest is set to $\sigmaru > 6$. This threshold is selected in particular for sources with few periods of increased activity during the whole observation time for HAWC.

For the case of Mrk 421, the object of interest for this study, we present the significance curve obtained with ARU for weekly periods in Figure~\ref{fig:aru_mrk421}. All the purple crosses show weekly periods where $\sigmaru$ is greater than 6, while the gray circles are all the weekly periods that are consistent with the steady emission.

We conclude this section by remarking that a key advantage of ARU is that there are no assumed spectral assumptions to derive the significance of transient sources. Furthermore, ARU also makes no assumptions about the different diffuse gamma-ray emissions, which introduces large systematic uncertainties when estimating the variability of sources with more traditional methods. Thus, ARU presents a methodology for use in conjunction with other analysis tools, e.g., \texttt{gammapy}\footnote{https://docs.gammapy.org/1.3/index.html}, in situations where spectral information for the transient source is available.

\section{Data Selection}
\label{sec:data}

We apply ARU to the 2460-transits dataset grouped (binned) in weekly and monthly time windows (periods) and get the values of $\sigmaru$ at the location of Mrk 421. We select 15 weekly periods where $\sigmaru > 6$ (see Figure \ref{fig:aru_mrk421}), and 9 monthly periods with $\sigmaru < 0$. The duration of each period is reported in Table~\ref{tab:timeinfo}. We associate the weekly periods as HES and the monthly periods as LES, based on the temporal coincidence with the reported on \cite{hawcMrk4212025}.

\begin{table}[t]
    \centering
    \begin{tabular}{c c c}
        \multicolumn{3}{c}{High Emission States} \\\hline\hline
        ID & $t_{\text{start}}$ & $t_{\text{stop}}$ \\
         0 & 57235.37         & 57246.58        \\ 
         1 & 58125.17         & 58135.79        \\ 
         2 & 58133.88         & 58142.88        \\ 
         3 & 58140.86         & 58151.27        \\ 
         4 & 58149.21         & 58160.94        \\ 
         5 & 58589.22         & 58598.65        \\ 
         6 & 59011.46         & 59021.88        \\ 
         7 & 59047.93         & 59060.07        \\ 
         8 & 59057.93         & 59069.88        \\ 
         9 & 59092.73         & 59102.10        \\ 
        10 & 59100.78         & 59110.93        \\ 
        11 & 59108.92         & 59119.91        \\ 
        12 & 59126.39         & 59137.22        \\ 
        13 & 59135.22         & 59145.97        \\ 
        14 & 59143.91         & 59154.02        \\ \hline
    \end{tabular}
    \begin{tabular}{c c c}
        \multicolumn{3}{c}{Low Emission States} \\\hline\hline
        ID & $t_{\text{start}}$ & $t_{\text{stop}}$ \\
         0 & 57355.19         & 57387.73        \\ 
         1 & 58009.82         & 58044.85        \\ 
         2 & 58401.83         & 58434.70        \\ 
         3 & 58609.13         & 58642.83        \\ 
         4 & 58735.06         & 58766.22        \\ 
         5 & 58908.42         & 58940.89        \\ 
         6 & 59156.04         & 59184.69        \\ 
         7 & 59357.11         & 59435.08        \\ 
         8 & 59829.96         & 59860.92        \\ \hline
    \end{tabular}
    \caption{Time information of Mrk 421 High Emission States and Low Emission States datasets used in this analysis. Time is expressed in Modified Julian Date (MJD).}
    \label{tab:timeinfo}
\end{table}

Once we identified the time duration of the different emission states of Mrk 421, we selected the data for a region of interest (ROI) of 8 degrees in radius, centered at the position of Mrk 421, and binned in energy. We use reconstructed data with the recent HAWC Pass 5 \citep{hawcPass5} for data selection.

The motivation for using monthly periods for the LES is that it guarantees sufficient signal-to-noise ratio to detect Mrk 421 and model its spectrum.
HES, on the other hand, are accumulated on weekly periods since this time frame provides enough counts in a shorter time, thanks to the enhanced source flux.

Our final sample results in $\thicksim270$ days and $\thicksim130$ days of data for LES and HES datasets, respectively.

\section{Modeling and Analysis}
\label{sec:anna}

As we mentioned in Section \ref{sec:aru}, ARU assists in selecting the short-term periods with a likely increase in activity. Then, we used \texttt{gammapy} \texttt{v 1.3} for the spectral analysis. More details on the analysis of HAWC data with \texttt{gammapy} can be found in \citep{hawc_gammapy}.

As the first step, we fitted only the spectrum of photons detected at Earth, i.e., the observed spectrum, by assuming an ECPL model. The cutoff energy ($E_{\text{c}}$) was fixed to $5~{\text{TeV}}$ for an easy comparison with the results presented in \citep{hawcMrk4212025}. The value of $5~{\text{TeV}}$ is also consistent with previous HAWC analyses with smaller datasets, see \citep{hawcMrk4212017,hawcMrk4212022}. The ECPL model is:

\begin{equation}\label{eq:obsflux_ecpl}
    \left.\frac{\mathrm{d}\phi}{\mathrm{d}E}\right|_{\text{obs}} = \phi_0\left(\frac{E}{1~{\text{TeV}}}\right)^{\alpha}\exp{\left(-\frac{E}{5~{\text{TeV}}}\right)},
\end{equation}
where $\alpha$ is the spectral index and $\phi_0$ is the normalization of the spectrum at the pivot energy. The pivot energy is fixed to  $1~\text{TeV}$ in this analysis. The observed flux is fit in all individual periods. We also obtained the joint spectrum for combined HES and LES, respectively. For each model tested, the joint spectrum was obtained from maximizing the total log-likelihood that results from adding the individual log-likelihood values.  

Secondly, we included the EBL attenuation as part of the spectrum modeling. Then, the observed flux is described by:

\begin{equation}\label{eq:obsspec_ebl_gral}
    \left.\frac{\mathrm{d}\phi}{\mathrm{d}E}\right|_{\text{obs}} = \left.\frac{\mathrm{d}\phi}{\mathrm{d}E}\right|_{\text{int}}\exp{\left(-\alpha_{\text{norm}}\tau_{\gamma\gamma}(E,z_{\text{Mrk 421}})\right)}.
\end{equation}
Here, the subscript int stands for intrinsic spectrum. The second term in Equation \ref{eq:obsspec_ebl_gral} is the attenuation factor induced by the interaction of TeV gamma rays with photons from the EBL radiation field, which is modeled by the opacity of the Universe to gamma rays $\tau_{\gamma\gamma}$, and the extra global normalization parameter $\alpha_{\text{norm}}$. We used four different EBL models to compute $\tau_{\gamma\gamma}$ and evaluate the systematic on the parameters of the intrinsic spectrum: Dominguez \citep{dominguez_2010}, Franceschini \citep{franceschini2008}, Finke \citep{Finke_2010}, and Saldana-Lopez \citep{saldana_2021}. Numerical tables of the different EBL models are available to use with \texttt{gammapy}. We define the nominal EBL scenarios when $\alpha_{\text{norm}}$ is equal to 1. The attenuation factor obtained using the nominal Dominguez model is our benchmark to present and discuss the main results of the analysis.

We tested different spectral hypotheses to describe the intrinsic spectrum. Our selection of spectral models is motivated by previous reports from MAGIC \citep{magicMrk4212007,magicMrk4212010} and VERITAS \citep{veritasMrk4212011}. The different spectral models tested are the SPL, LP, and ECPL. Such models are given by:

\begin{align}
    \left.\frac{\mathrm{d}\phi}{\mathrm{d}E}\right|_{\text{int}}^{\text{SPL}} & = \phi_0\left(\frac{E}{1~{\text{TeV}}}\right)^{\alpha}, \label{eq:int_spl}\\
    \left.\frac{\mathrm{d}\phi}{\mathrm{d}E}\right|_{\text{int}}^{\text{LP}} & = \phi_0\left(\frac{E}{1~{\text{TeV}}}\right)^{\alpha-\beta\ln\left(\frac{E}{1~{\text{TeV}}}\right)}, \label{eq:int_lp}\\
    \left.\frac{\mathrm{d}\phi}{\mathrm{d}E}\right|_{\text{int}}^{\text{ECPL}} & = \phi_0\left(\frac{E}{1~{\text{TeV}}}\right)^{\alpha}\exp{\left(-\lambda~E\right)}, \label{eq:int_ecpl}
\end{align}
with $\alpha$ the spectral index, $\beta$ the curvature index, and $\lambda$ the inverse of the cutoff energy. For all the models, $\phi_0$ is the normalization of the spectrum. For the ECPL model we tested two cases when $\lambda$ was fixed to $0.2~\text{TeV}^{-1}$ (or $E_\text{c}$ fixed to $5~\text{TeV}$), and $\lambda$ free to float during the fit. During this step, we tested for linear and quadratic correlation of the spectral index and the photon flux $\Phi$ (integrated for energies between $1~\text{TeV}$ and $30~\text{TeV}$). Additionally, we used a Maximum Likelihood (ML) Linear Fit to test for hardening of the spectrum as a function of the photon flux $\Phi$.

For the joint spectrum derived with the LES dataset, we only estimate the impact of the EBL on the parameters of the intrinsic spectrum. As previously stated, LES have less constraining power for $\nu I_\nu$ as the flux is too low, resulting in larger uncertainties in the spectrum.

\begin{figure}[ht]
    \centering
    \includegraphics[width=0.75\linewidth]{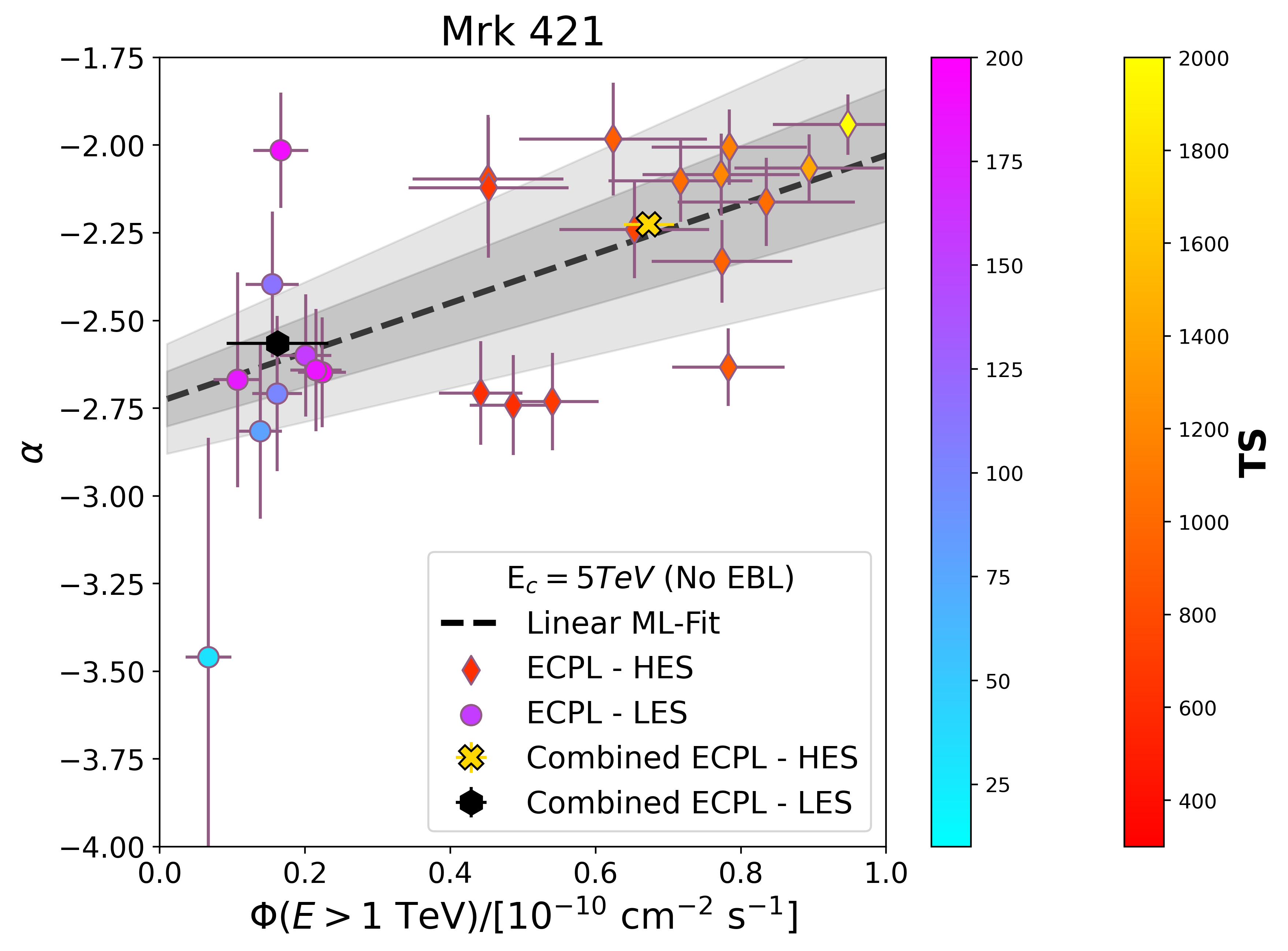}
    \caption{Scatter of weekly HES (diamonds) and monthly LES (circles) periods in the ($\alpha$--$\Phi$) parameter space. Parameters obtained for the observed spectrum test only. The color bars indicate the $TS$. The cross and the hexagon show the results for the joint fits of HES and LES datasets respectively. We also include the best ML linear fit (black dashed line) and the $1\sigma$ and $2\sigma$ uncertainties bands (shaded gray bands) around the linear fit.}
    \label{fig:2Dscatter_obs}
\end{figure}

We focus now on the joint spectrum obtained from the HES dataset. The third step consisted of estimating the impact of the EBL on the parameters of the intrinsic spectrum. Here, we also determined the best model that describes the intrinsic spectrum and estimated the significance with respect to the SPL case. Once we constrained the intrinsic spectrum, we freed the extra parameter $\alpha_\text{norm}$ for the 4 different EBL models and estimate the change in $TS$ with respect to the nominal ($\alpha_\text{norm} = 1$) cases. The upper limits (95\% C.L.) on $\alpha_\text{norm}$ are computed as the maximum value of $\alpha_\text{norm}$ where $\sqrt{TS} = 2$. Because $\tau_{\gamma\gamma}$ is proportional to $\nu I_\nu$ \citep{Biteau_2015} we can define $\tau'_{\gamma\gamma} = \alpha_\text{norm}\times\tau_{\gamma\gamma} \propto\nu I_\nu$. Combined with the fact that $\alpha_\text{norm}$ is independent of the TeV gamma-ray and EBL photon energies, we can directly convert upper limits on $\alpha_\text{norm}$ to upper limits on $\nu I_\nu$. More details can be also found in \citepalias{hessM87}.

\section{Results}
\label{sec:results}

In this section, we present the results obtained. For the 2D scatter plots, we use two different color bars to indicate whether a point belongs to the HES or LES datasets. The color bars
indicate either test statistic\footnote{This $TS$ is obtained from the usual Maximum Likelihood Estimation of model parameters} $TS$ or photon flux $\Phi$.

\subsection{Observed Flux}

Figure \ref{fig:2Dscatter_obs} shows the scatter of the HES (warm color diamonds) and LES (cool color circles) periods in the $(\alpha,\Phi)$ parameter space. The color bars refer to the TS obtained from the fit. First, we note as expected, that HES and LES periods are separated in terms of the photon flux $\Phi$, with HES periods having fluxes above $0.4\times10^{-10}~\text{cm}^{-2}~\text{s}^{-1}$. We also found a moderate linear  correlation ($r$) with a value of $0.63~[p=1.01\times10^{-3}]$, suggesting a hardening of the spectrum for periods of increased activity of Mrk 421. The black dashed line in Fig. \ref{fig:2Dscatter_obs} shows the results of the best linear fit, and the gray bands indicate the $1\sigma$ and $2\sigma$ uncertainty bands of the linear fit. The gold cross and black hexagon represent the results of the joint fit for each dataset. We find a clear distinction between the spectra, with the joint LES spectra having a softer spectrum than their counterpart during the HES periods.

\subsection{Intrinsic Spectrum}
\label{sec:intrinsic}

In the following, we present the results obtained for the different intrinsic spectral models tested using nominal ($\alpha_\text{norm} = 1$) EBL models. We focus primarily on the results for the SPL and ECPL models. The LP model is not shown as it offers no improvement in TS over the SPL model ($\Delta TS\leq9$, see Table \ref{tab:delta_TS}).

\begin{figure}[ht]
    \centering
    \includegraphics[width=0.75\linewidth]{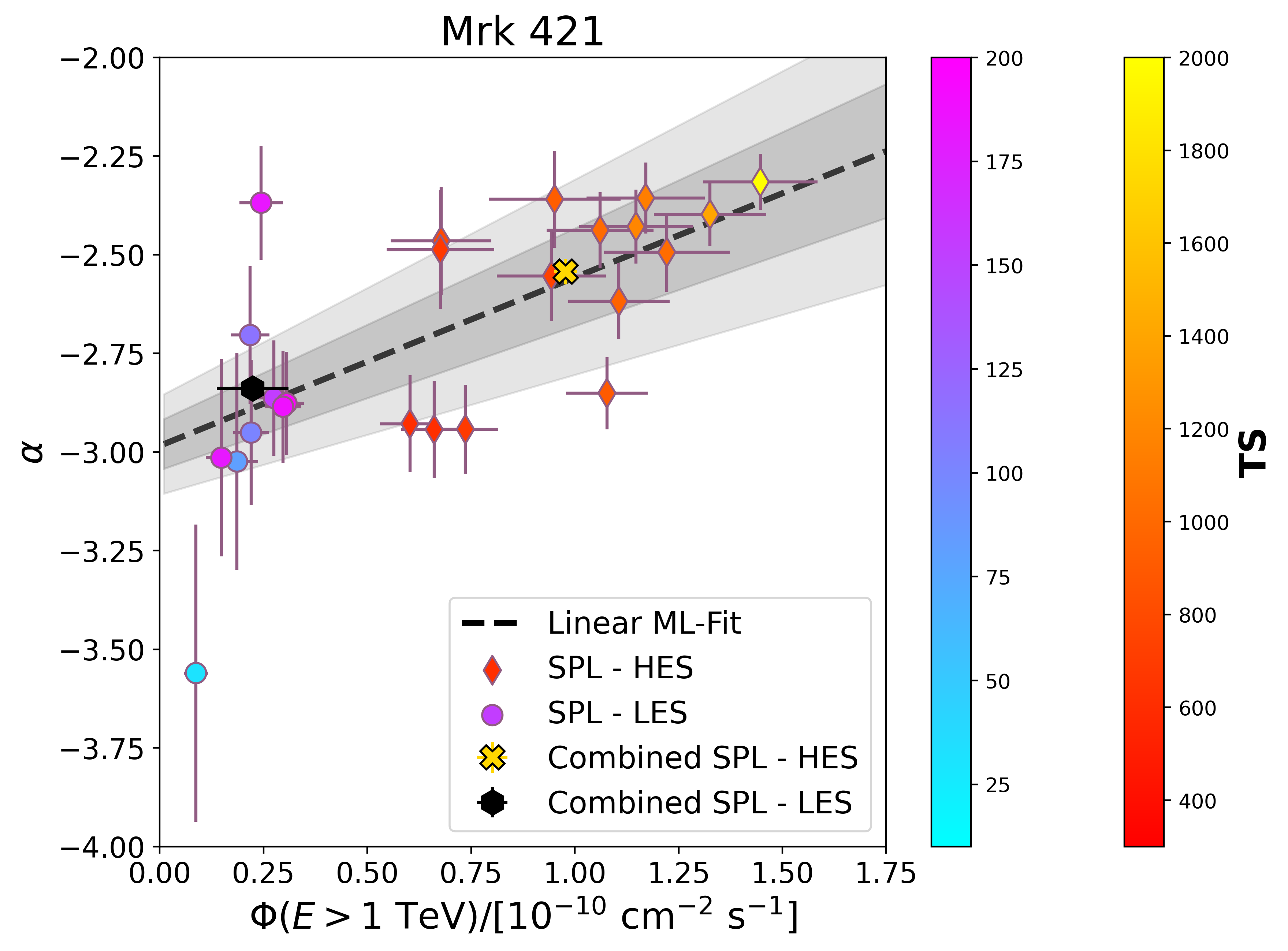}
    \includegraphics[width=0.75\linewidth]{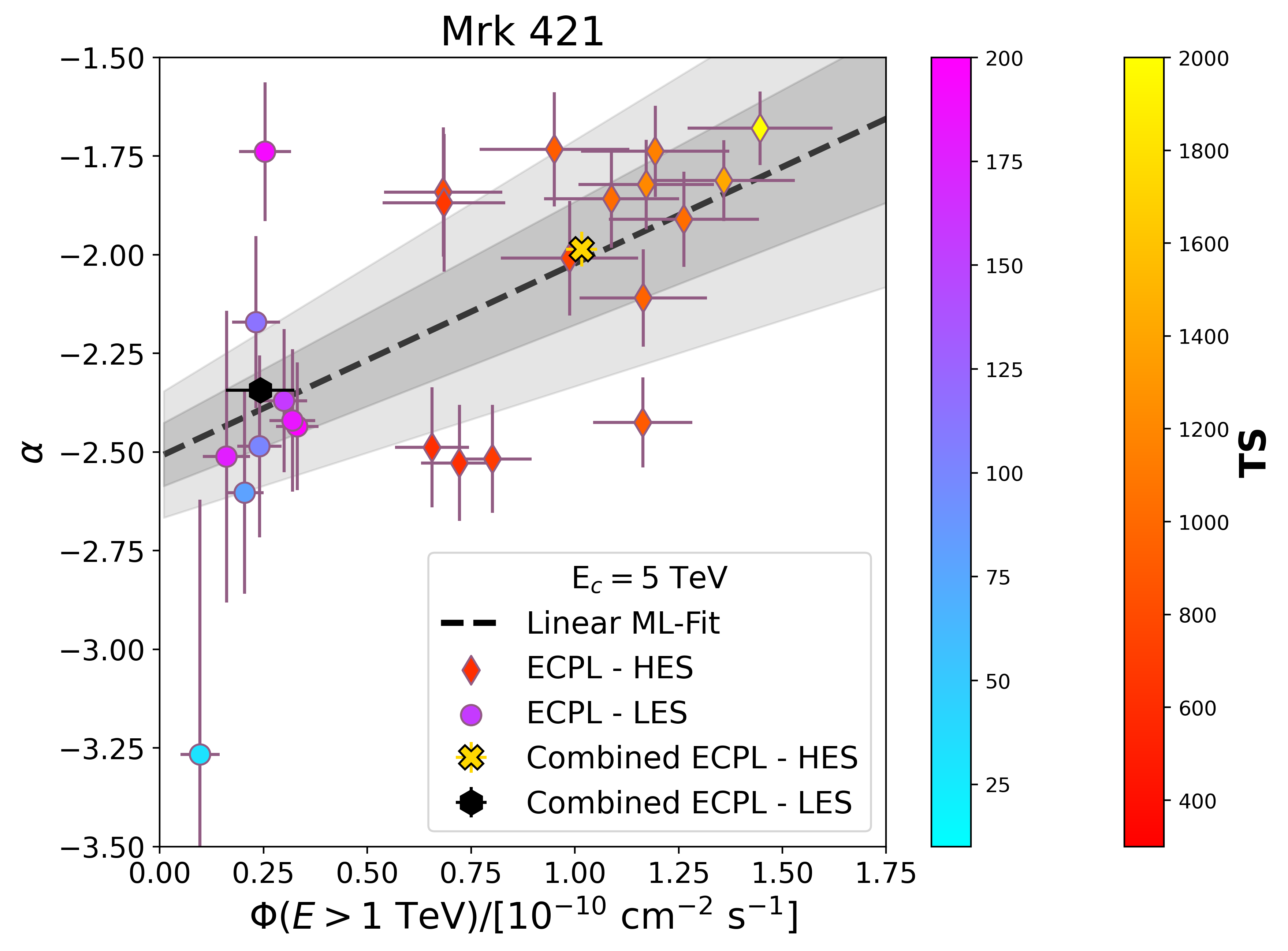}
    \caption{Scatter of weekly HES (diamonds) and monthly LES (circles) periods in the ($\alpha$--$\Phi$) parameter space. Parameters obtained for the intrinsic spectrum assuming a SPL model (upper panel) and ECPL model (lower panel). The color bars indicate the $TS$. The cross and the  hexagon show the results for the joint fits of HES and LES datasets respectively. We also include the best ML linear fit (black dashed line) and the $1\sigma$ and $2\sigma$ uncertainties bands (shaded gray bands) around the linear fit.}
    \label{fig:2Dscatter_spl_ecpl}
\end{figure}

We show the results for the SPL and ECPL (with $\lambda$ fixed to $0.2~\text{TeV}^{-1}$) in Figure~\ref{fig:2Dscatter_spl_ecpl}. Similar to the case of the observed flux (Figure \ref{fig:2Dscatter_obs}), we note that HES and LES periods show a clear separation in terms of the photon flux. We also obtained a moderate linear correlation for both spectral models with $r=0.69~[p=1.92\times10^{-4}]$ for SPL and $r=0.64~[p=8.24\times10^{-4}]$ for ECPL (fixed $\lambda$), respectively. The hardening of the spectrum is also supported by the linear fit (black dashed line) shown in both plots of Figure \ref{fig:2Dscatter_spl_ecpl}. The joint spectra for HES (gold cross) and LES (black hexagon) have a separation of $\thicksim0.3$ in the spectral index, similar to the observed spectra.

\begin{figure}[htb]
    \centering
    \includegraphics[width=0.55\linewidth]{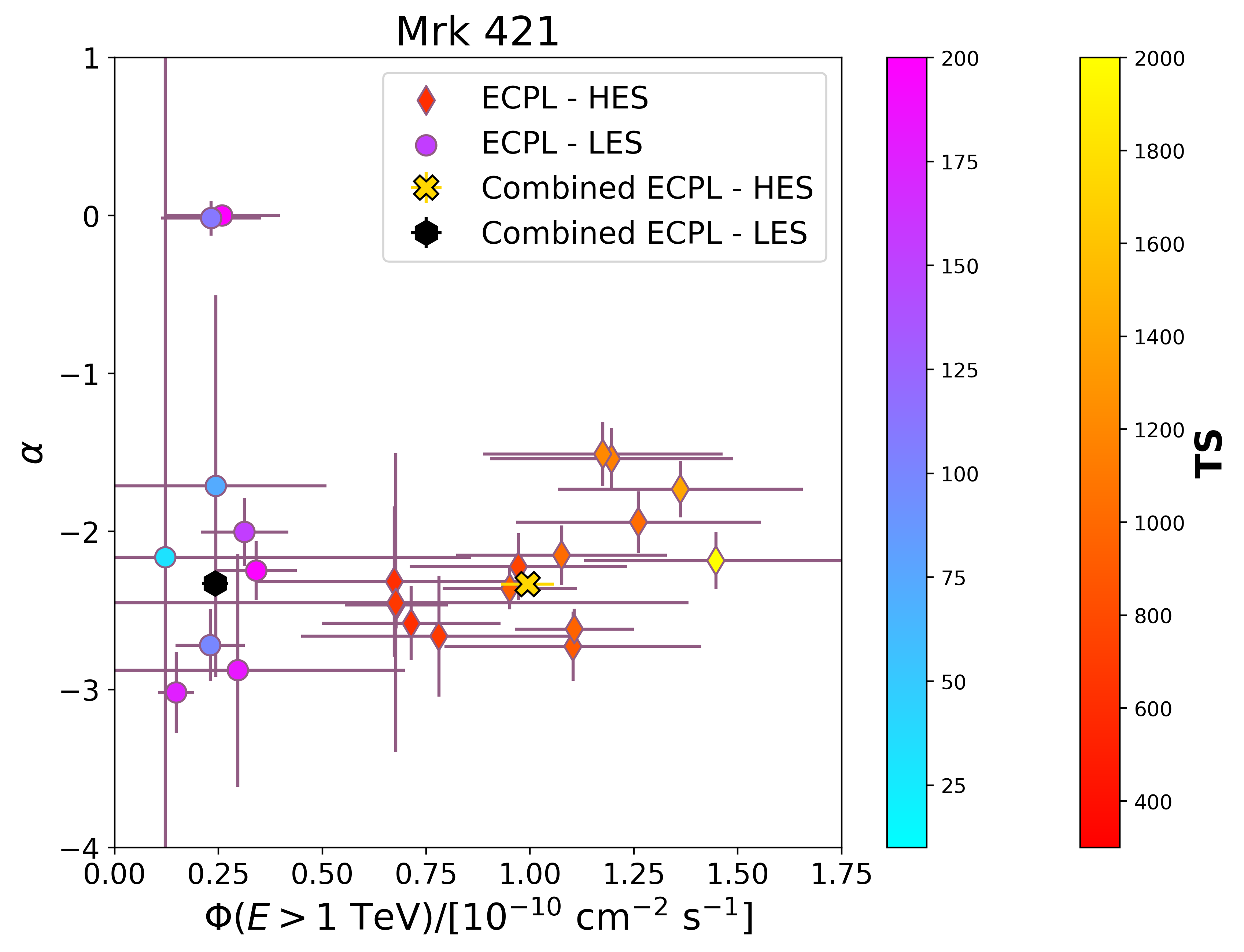}
    \includegraphics[width=0.55\linewidth]{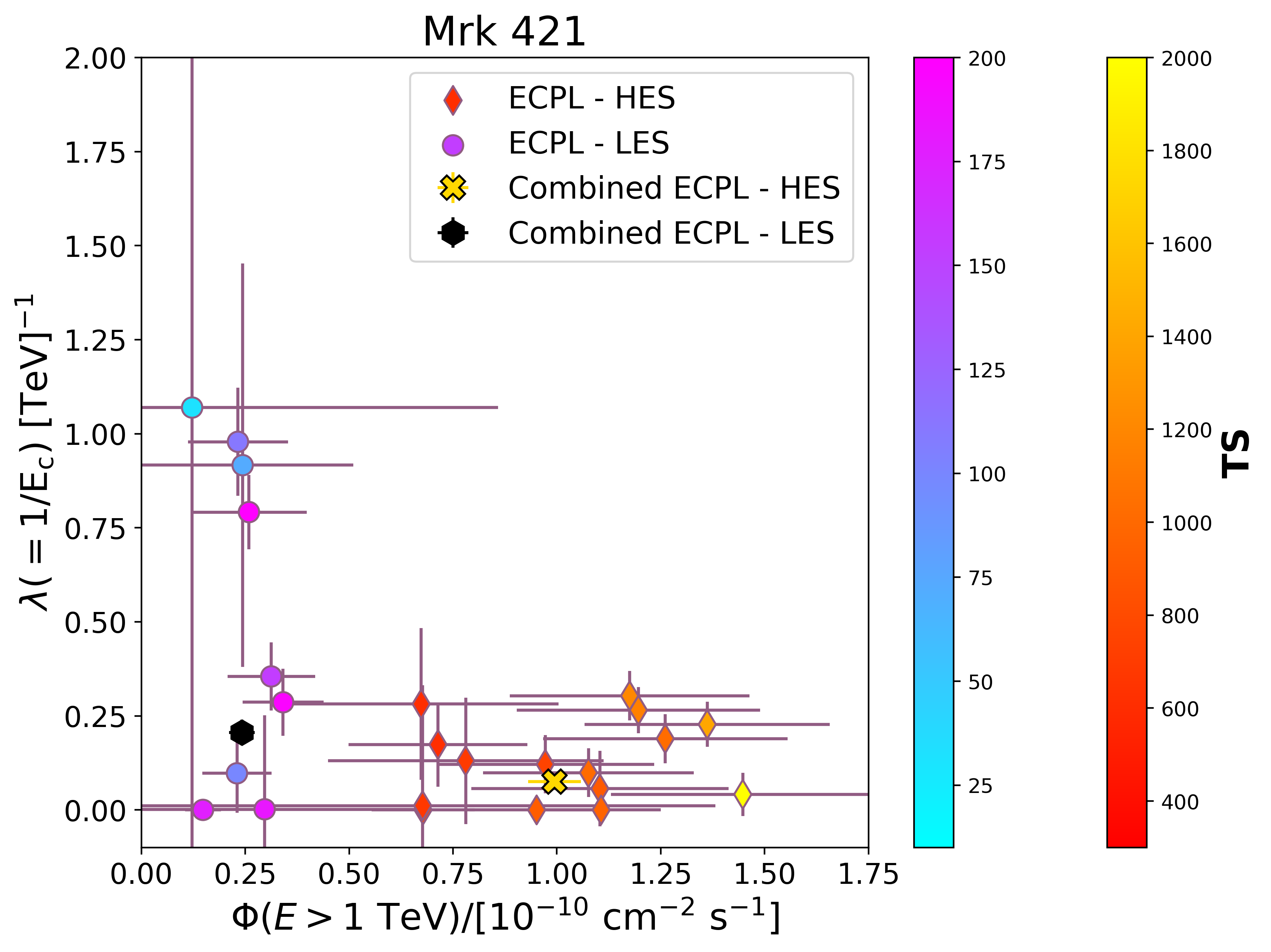}
    \includegraphics[width=0.55\linewidth]{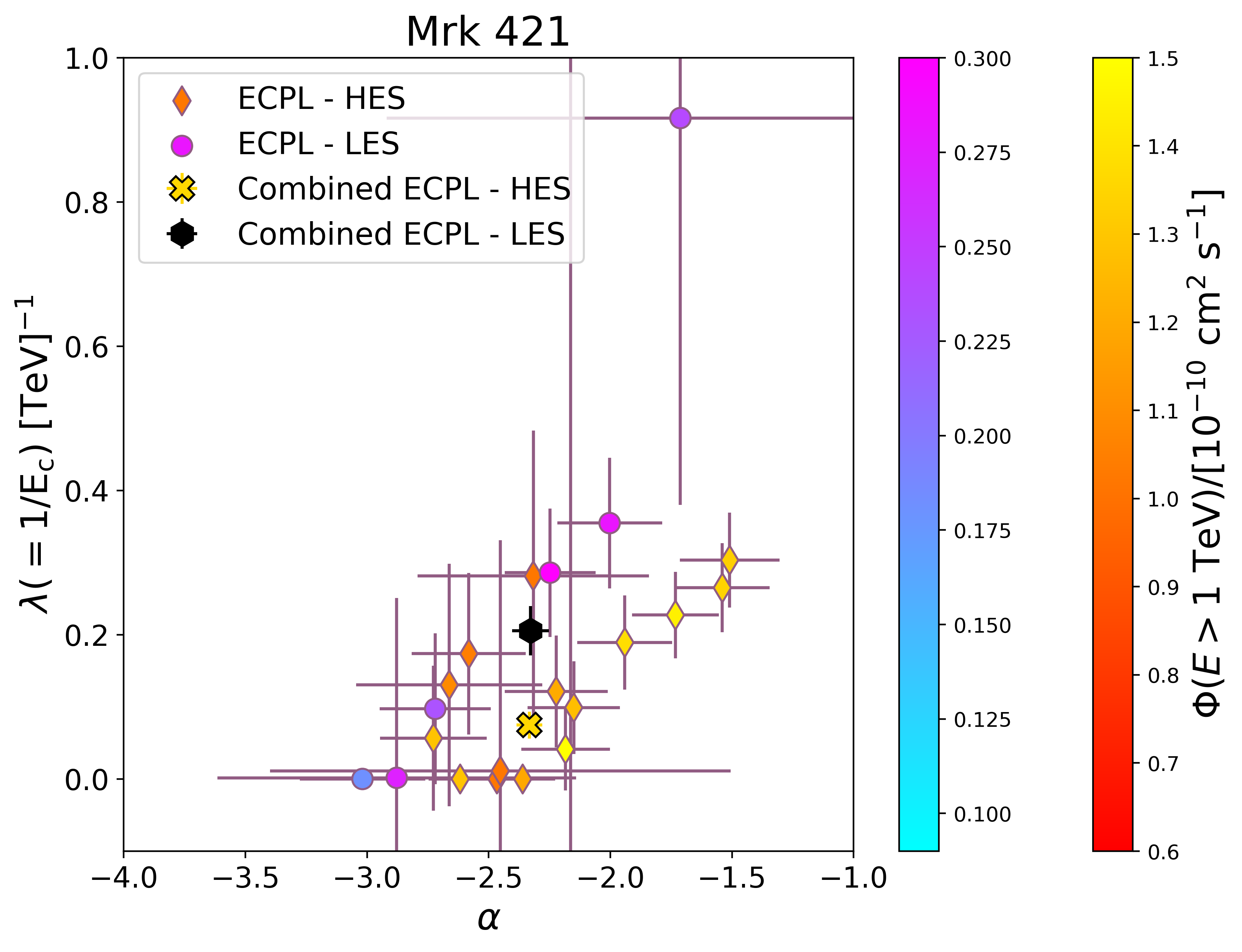}
    \caption{Scatter of weekly HES (diamonds) and monthly LES (circles) periods assuming an ECPL model. The different parameter spaces are ($\alpha$--$\Phi$) (upper panel), ($\lambda$--$\Phi$) (middle panel) and ($\lambda$--$\alpha$) (lower panel). The color bars indicate the $TS$ (left and middle panels) and photon flux $\Phi$ (right panel). The cross and the hexagon show the results for the joint fits of HES and LES datasets respectively.}
    \label{fig:2Dscatter_ecpl}
\end{figure}

On the other hand, the results obtained when $\lambda$ is free in the ECPL model exhibit a more random behavior (see Figure~\ref{fig:2Dscatter_ecpl}). The markers and colors are the same as in Figures~\ref{fig:2Dscatter_obs} and \ref{fig:2Dscatter_spl_ecpl}. We note that the clear separation in photon flux between the LES and HES periods remains. However, the uncertainties in the photon fluxes are larger in comparison to the SPL scenario. The upper panel of Figure \ref{fig:2Dscatter_ecpl} reveals that there is no apparent hardening of the spectrum as the activity of Mrk 421 increases. This is also obtained for the joint spectra of HES and LES datasets that have the same spectral index under this scenario. The middle and lower panels of Figure \ref{fig:2Dscatter_ecpl} show the position of the different HES and LES periods in the parameter spaces of ($\lambda$--$\Phi$) and ($\lambda$--$\alpha$) respectively. This is done to investigate if a change in the inverse of the cutoff energy $\lambda$ correlates with changes of the other two parameters. Qualitatively, an apparent shift to smaller values of $\lambda$ as a function of the photon flux $\Phi$ could be observed in the scatter of individual periods. However, because of the uncertainties, it is difficult to claim any particular trend between $\lambda$ and the other two parameters in the ECPL model. Joint spectra of HES and LES periods show different values of $\lambda$ ($\lambda_\text{HES} = 0.07 \pm 0.02~\text{TeV}^{-1}$ and $\lambda_\text{LES} = 0.21\pm0.03~\text{TeV}^{-1}$) as observed in the middle and right panels of Figure \ref{fig:2Dscatter_ecpl}. This translates to a shift of the cutoff to higher energies during periods of increased activity of Mrk 421.

\begin{table}[ht]
    \centering
    \begin{tabular}{c||c|c}
       Model  & $\Delta TS$ (HES) & $\Delta TS$ (LES) \\\hline\hline
       LP     &        7          &        9          \\\hline
       ECPL   &        15         &        11         \\\hline
       ECPL ($\lambda^\text{fix}$)   &        -12        &        11        \\
    \end{tabular}
    \caption{Difference in TS ($\Delta TS$) of the different models with respect to SPL for the joint HES and LES spectra.}
    \label{tab:delta_TS}
\end{table}

As seen in Table \ref{tab:delta_TS}, the ECPL model with $\lambda$ free has an increment in $TS$ (with respect to the SPL spectrum) of 15 to explain the spectrum combining all the periods of the HES dataset.
For LES periods, the best fit to the joint spectrum is obtained for the ECPL model, regardless of whether $\lambda$ is free during the fit or not. We estimated the cutoff energy for HES joint spectrum at $13 \pm 3$\,TeV and a significance of $3.8\sigma$. Similarly, the cutoff for the LES joint spectrum is $4.9\pm 0.8$\,TeV with a significance of $3.3\sigma$. The values of the cutoff found in the intrinsic spectra of Mrk 421 are different from the expected cutoff induced by the EBL. For example, for the Dominguez model, the expected cutoff is around $7.2~\text{TeV}$.

\begin{figure}[ht]
    \centering
    \includegraphics[width=0.75\linewidth]{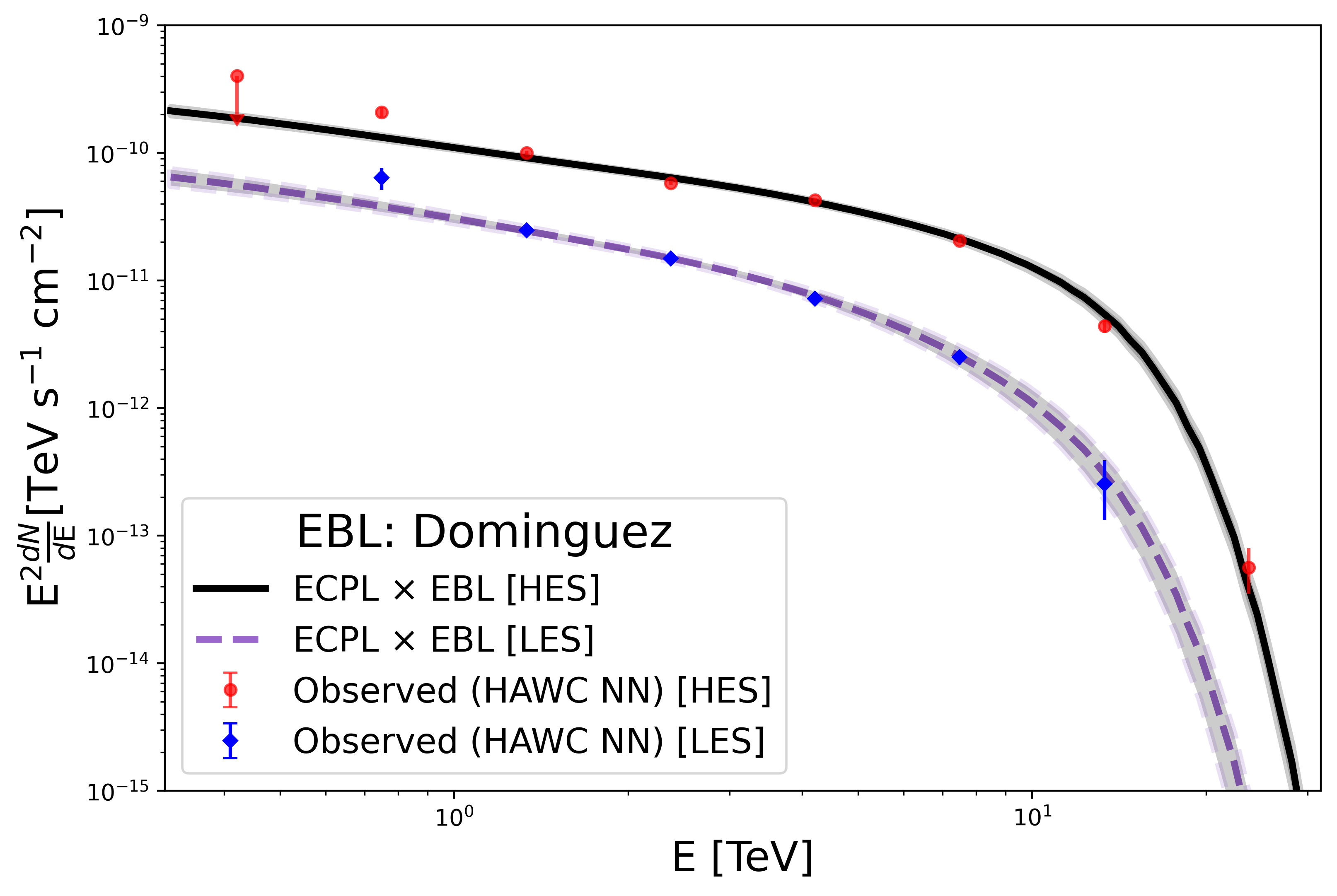}
    \caption{Spectrum of Mrk 421 for two different emission states. Intrinsic joint spectrum obtained for HES (red circles) and LES (blue diamonds) datasets. The minimum $TS$ value to consider a flux point is $TS = 4$. For the HES spectrum, the last point has a significance of $\sqrt{TS} = 3.14$. The black solid and purple dashed lines represent the best ECPL model obtained after fitting. The bands correspond to $1\sigma$ statistical uncertainty after error propagation.}
    \label{fig:mrk421_SED_ecpl}
\end{figure}

Figure \ref{fig:mrk421_SED_ecpl} shows the SED of Mrk 421 for HES (red circles) and LES (blue diamonds) datasets. We observe the clear difference in intensity between both periods. The dashed lines correspond to the best fit ECPL model. We also observe that the joint spectrum of HES periods has a flux point at energies centered around $24~\text{TeV}$ and a significance of $3.14\sigma$, evidencing the emission from Mrk 421 at energies above $18~\text{TeV}$. An additional sample of HES periods can, eventually, lead to the detection of emission of Mrk 421 at such high energies. On the other hand, the energy of the lower end in the spectrum corresponds to $0.42~\text{TeV}$, where we were only able to report an upper limit to the flux due to low statistics.

\subsection{Constraints on the specific intensity of EBL}

\begin{figure}[ht]
    \centering
    \includegraphics[width=0.45\linewidth]{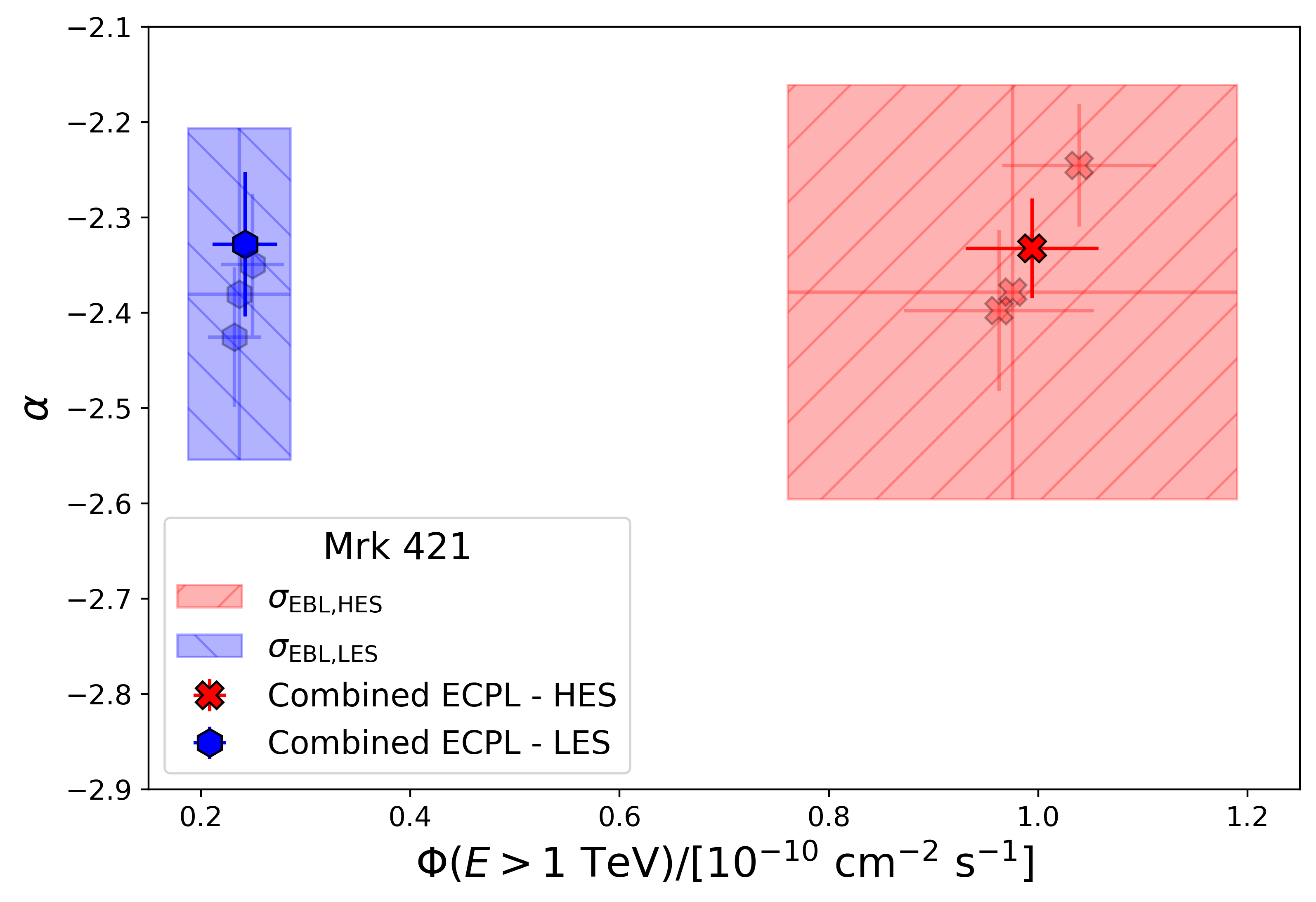}
    \includegraphics[width=0.45\linewidth]{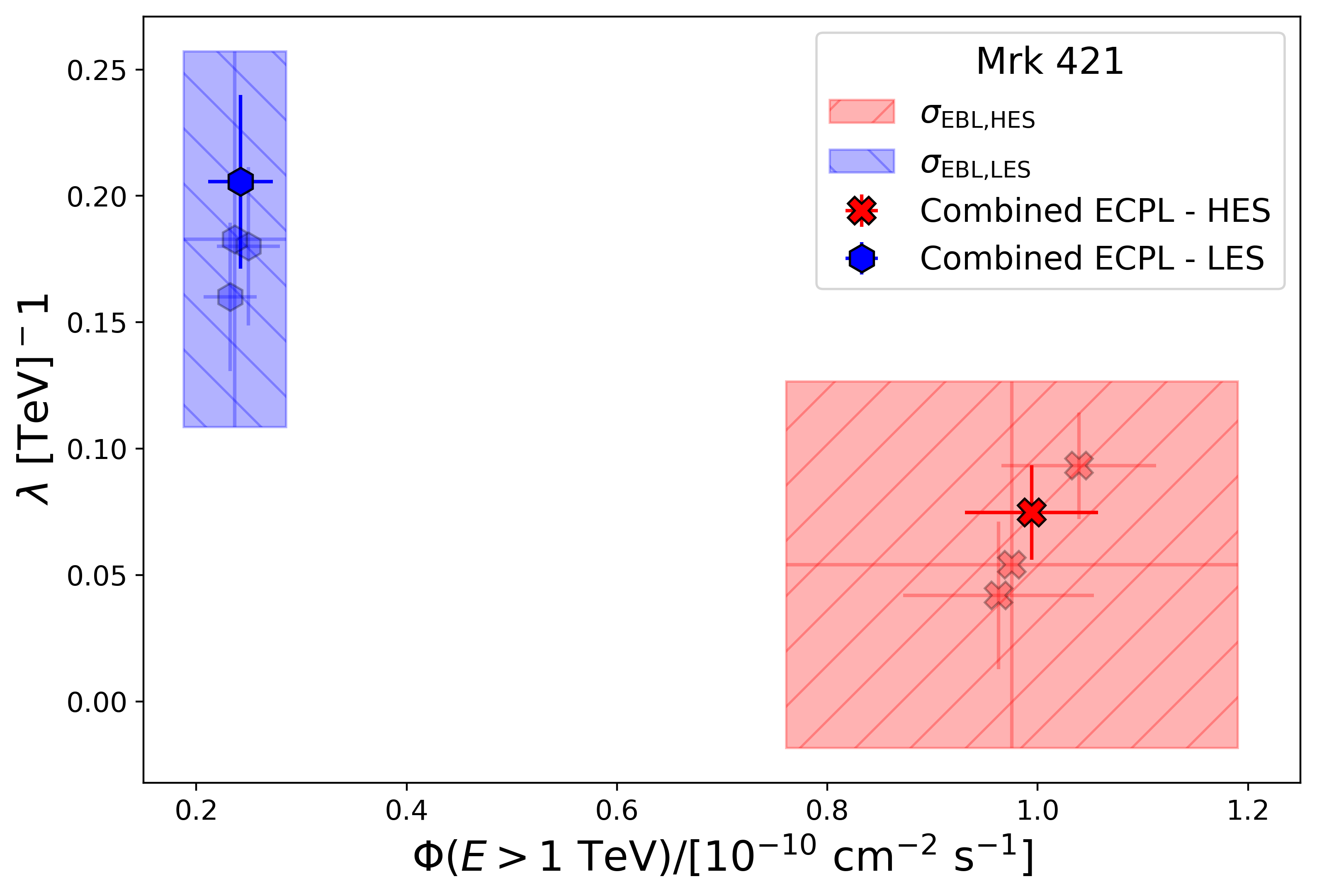}
    \caption{Impact of the EBL modeling on the combined spectrum of HES (red crosses) and LES (blue hexagons) datasets. \textit{Left}: Spectral index versus photon flux ($\alpha$--$\Phi$) shows no variation for spectral index for different activity periods. \textit{Right}: Inverse cutoff energy versus integrated flux ($\lambda-\Phi$) exhibits shift towards smaller values. In both cases, the flux is integrated for gamma-ray energies above 1 TeV and expressed in units of $10^{-10}$ cm$^{-2}$\,s$^{-1}$. The solid markers are the results for our benchmark model (nominal Dominguez). The uncertainty boxes were computed with the maximal variation from the four EBL models tested.}
    \label{fig:ecpl_ebl_variation}
\end{figure}

Before presenting the resulting constraints on the EBL specific intensity $\nu I_\nu$, we show the impact of the EBL modeling on the estimation of the parameters for the best fit model. Figure \ref{fig:ecpl_ebl_variation} shows the uncertainty box obtained for the 4 different nominal models (see Section \ref{sec:anna}). The solid markers in both panels of Figure \ref{fig:ecpl_ebl_variation} correspond to the results for our benchmark model. The uncertainty boxes are constructed by only taking the maximum (minimum) variation of the values $\{p+\Delta p\}$ ($\{p-\Delta p\}$), with $p$ any parameter of the ECPL model and $\Delta p$ its uncertainty. The uncertainty boxes do not represent any confidence interval and only show the possible variation of the parameters when a different EBL model is selected. The upper panel in Figure \ref{fig:ecpl_ebl_variation} indicates that the spectral index shows no significant variation between high and low activity periods, whereas the inverse of the cutoff energy $\lambda$ is consistent with lower values during periods of increased source flux  (see lower panel of Figure \ref{fig:ecpl_ebl_variation}). The larger uncertainty in the HES and LES datasets is obtained when the Saldana-Lopez model is used to estimate the EBL attenuation factor. However, our conclusion about the ECPL does not change when considering this particular model.

\begin{figure}[ht]
    \centering
    \includegraphics[width=0.75\linewidth]{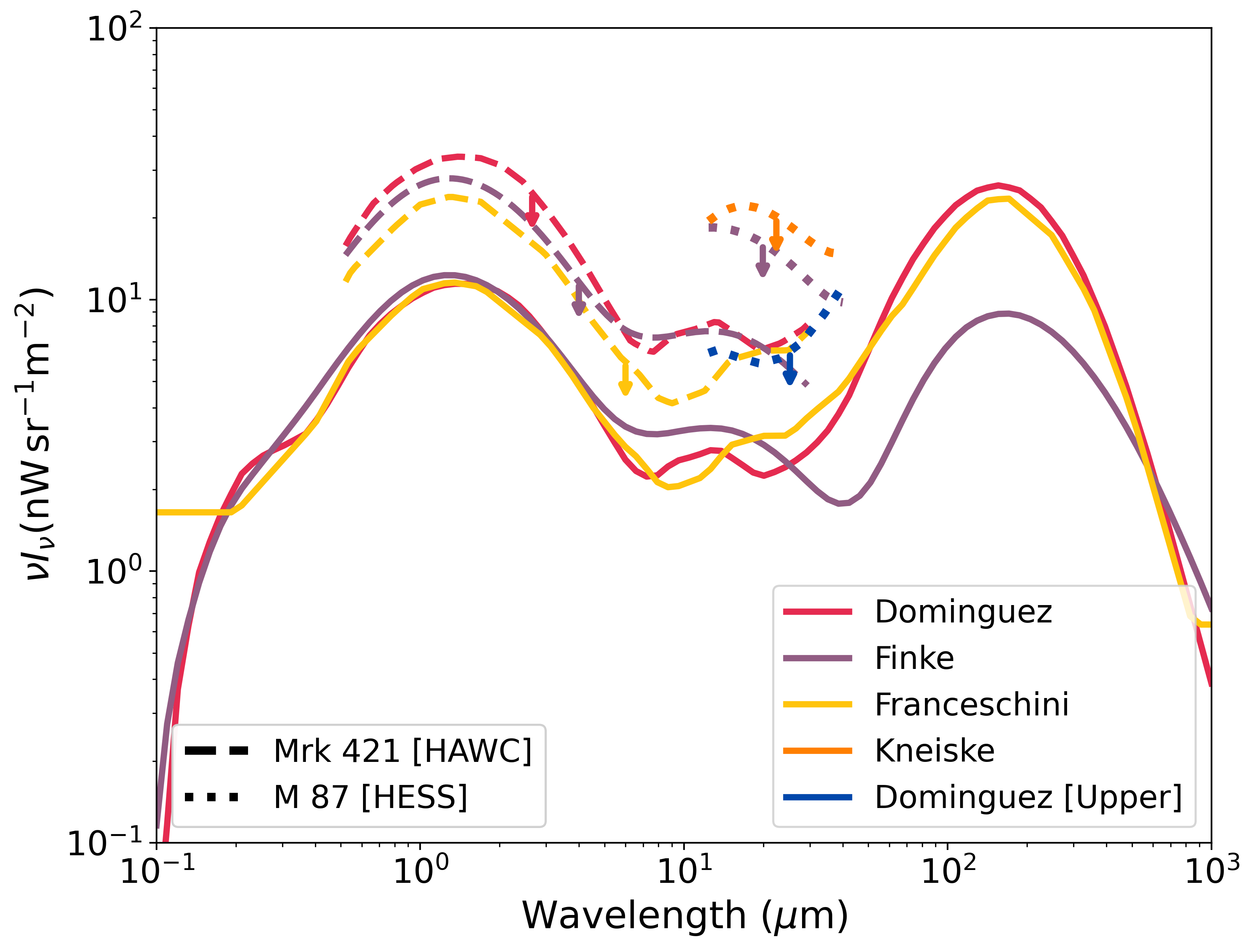}
    \caption{Upper Limits to EBL specific intensity ($\nu I_\nu$) as a function of photon wavelength for different EBL models. The upper limits are derived from the best fit to the intrinsic spectrum and finding the 95\% C.L. to $\alpha_\text{norm}$. Dashed lines represent the upper limits obtained from our analysis of Mrk 421 during HES periods. Dotted lines are the results obtained by H.E.S.S. observations of M 87 during the flaring periods \citepalias{hessM87}. Solid lines are the estimated specific intensity for the different EBL models used in our analysis and by H.E.S.S.}
    \label{fig:ebl_uls}
\end{figure}

Figure \ref{fig:ebl_uls} presents the upper limits obtained from the joint HES periods on the specific intensity $\nu I_\nu$ (95\% C.L.) derived for all the  EBL models, except Saldana-Lopez because of the large uncertainties found during the fit of the nominal case which prevented proper convergence on the parameter space. The upper limits on $\nu I_\nu$ were derived in the wavelength interval from 0.52 to 29.40 $\mu\text{m}$, corresponding to the energy range from $0.42~\text{TeV}$ to $24~\text{TeV}$ where the spectral fit was done, see Section \ref{sec:intrinsic} and Figure \ref{fig:mrk421_SED_ecpl}. We use the expression $\lambda_\text{EBL}=1.24\times\left(E_\gamma/1~\text{TeV}\right)$ to convert between gamma-ray energies and the value of wavelength of the EBL photons \citep{franceschini_IFEBL}. We also present the upper limits (dotted lines in Figure \ref{fig:ebl_uls}) obtained by the H.E.S.S. collaboration using observations of M 87 during flaring periods in the energy range from $5~\text{TeV}$ to $32~\text{TeV}$ \citep{hessM87}. Our upper limits for HES of Mrk~421 are consistent with those reported by H.E.S.S. for M~87 in high state.

\section{Summary and Discussion}
\label{sec:discussion}

\subsection{Spectral Curvature, Cutoff, and EBL Constraints}

Using ARU, we selected periods of increased activity for Mrk 421 ($\sigmaru>6$) at weekly scales (see Figure~\ref{fig:aru_mrk421}) and estimated the best-fit spectrum for individual periods and the joint dataset. We performed the same analysis for a sample of 9 monthly periods with low emission ($\sigmaru < 0$) to investigate if there are differences between periods of increased and low activity. The spectral analysis shows a hardening of the spectrum (moderate correlation of $r = 0.69$) when a SPL describes the spectrum (see upper panel of Figure~\ref{fig:2Dscatter_spl_ecpl}). However, the best-fit model of the intrinsic spectrum for the joint datasets is an ECPL model. The cutoff energies were estimated at $13\pm3~\text{TeV}$ ($3.8~\sigma$) and $4.9\pm0.8~\text{TeV}$ ($3.3~\sigma$), for HES and LES periods, respectively. Using the results obtained for the ECPL spectrum of the joint HES periods, we estimated upper limits (at 95\% C.L.) on the specific intensity of EBL photons $\nu I_\nu$ (see Figure~\ref{fig:ebl_uls}).

The hardening of the spectrum is also observed when the intrinsic spectrum is assumed to be an ECPL ($\lambda$ fixed to $0.2~\text{TeV}^{-1}$, see the lower panel of Figure \ref{fig:2Dscatter_spl_ecpl}). While the correlation is moderate,
this is consistent with observations by MAGIC \citep{magicMrk4212007,magicMrk4212010,magicMrk4212012} and VERITAS \citep{veritasMrk4212011}. We do not observe any saturation/flattening on the hardening of the spectrum as presented in \citep{mrk421_20152016_low_emission} where the authors report that the spectral index reach a constant value during periods of very high emission in the TeV band. The lack of evidence on the flattening is probably because of our limited sample (24 observation periods) against the $\thicksim240$ daily observations from MAGIC and FACT.

However, the linear correlation disappears when we free $\lambda$ during the fit to test if an ECPL model better describes the intrinsic spectra of individual periods. While the LES and HES joint spectra is better explained by ECPL models, no evidence of a cutoff was observed in the individual periods.

Using a smaller dataset, HAWC reported that the long-term spectrum of Mrk 421 is well described by an ECPL model with cutoff energy at $5.1\pm1.6~\text{TeV}$ \citep{hawcMrk4212022}. The Gilmore model was used in \citep{hawcMrk4212022} to estimate the EBL attenuation factor. While models compute the EBL opacity $\tau_{\gamma\gamma}$ differently, the cutoff energy in our analysis ($4.9\pm0.8~\text{TeV}$) and \citep{hawcMrk4212022} are consistent between them.

As pointed out by the results of previous observation campaigns, the appearance of the cutoff in the joint spectra of low-emission and high-emission states indicates that the cutoff is intrinsic to the source as opposed to being produced by EBL-photon interactions.  This notion is supported further by the discrepancy between the values of the cutoff energies for both emission states and the expected cutoff induced by the EBL. For the different EBL models used in this study, the cutoff is in the range from 6.6 to $\thicksim7.2~\text{TeV}$; while the cutoff energies for LES and HES are $4.9\pm0.8~\text{TeV}$ and $13\pm3~\text{TeV}$, respectively. One possible scenario to explain the observed cutoff at TeV would originate from the interaction with a dense radiation field at infrared (IR) wavelengths produced in external regions close to the central engine. For example, observations from the Mount Abu Infrared Telescope \citep{nearIRMrk421} show evidence of variability of the IR emission in Mrk 421 that can be explained from a contribution of the accretion disk. The authors in \citep{irpropertiesblazars} argue that the superposition of the components from the jet, the host galaxy, and a contribution from dust can explain IR emission. These two previous results support the scenario of an IR external radiation field that can enter into the gamma-ray emission region and induce an extra attenuation factor. However, the exact dynamics, other interaction terms, and the exact emission region of the external photons need to be evaluated more precisely.

\begin{figure}[ht]
    \centering
    \includegraphics[width=0.85\linewidth]{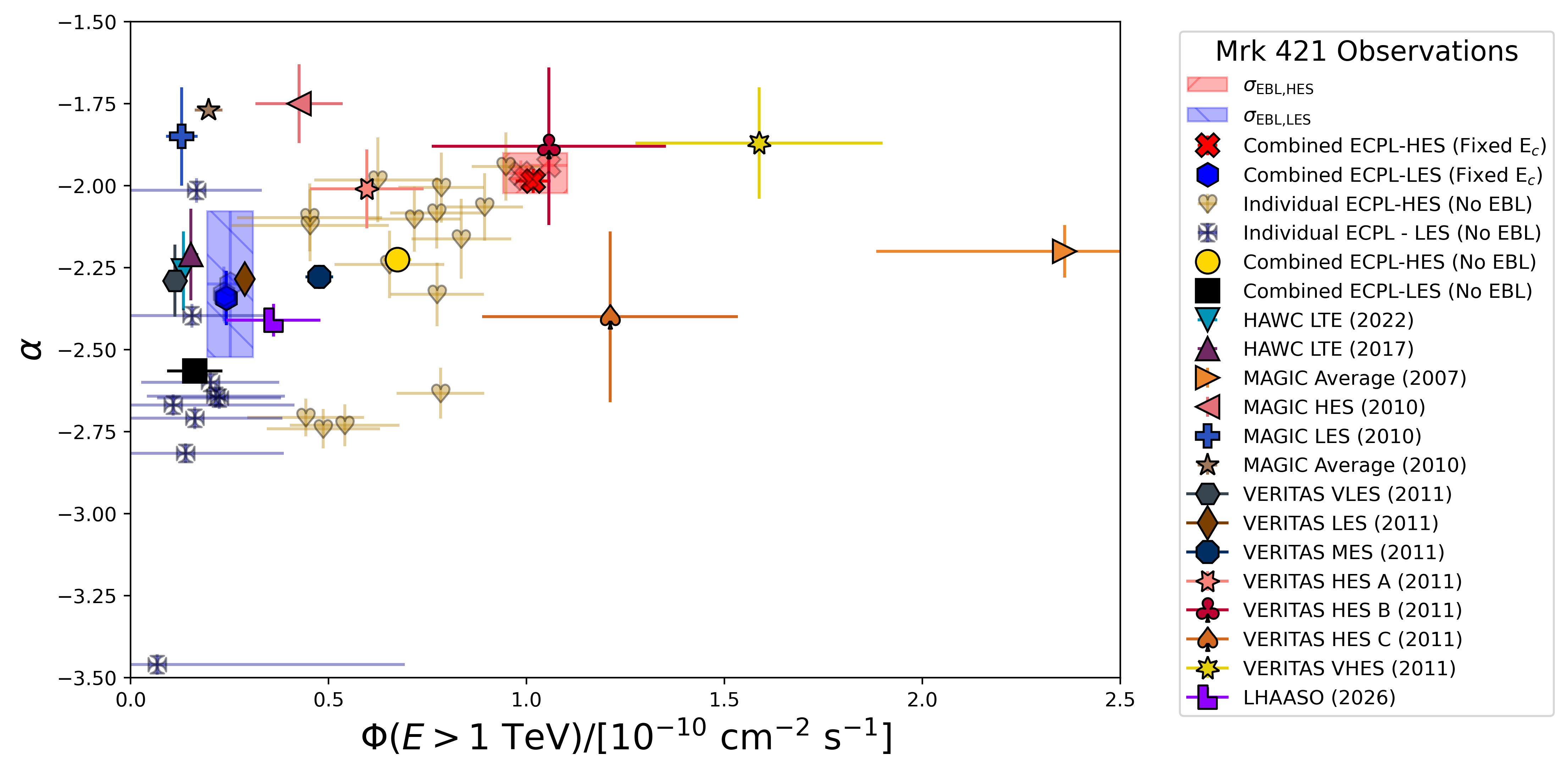}
    \includegraphics[width=0.85\linewidth]{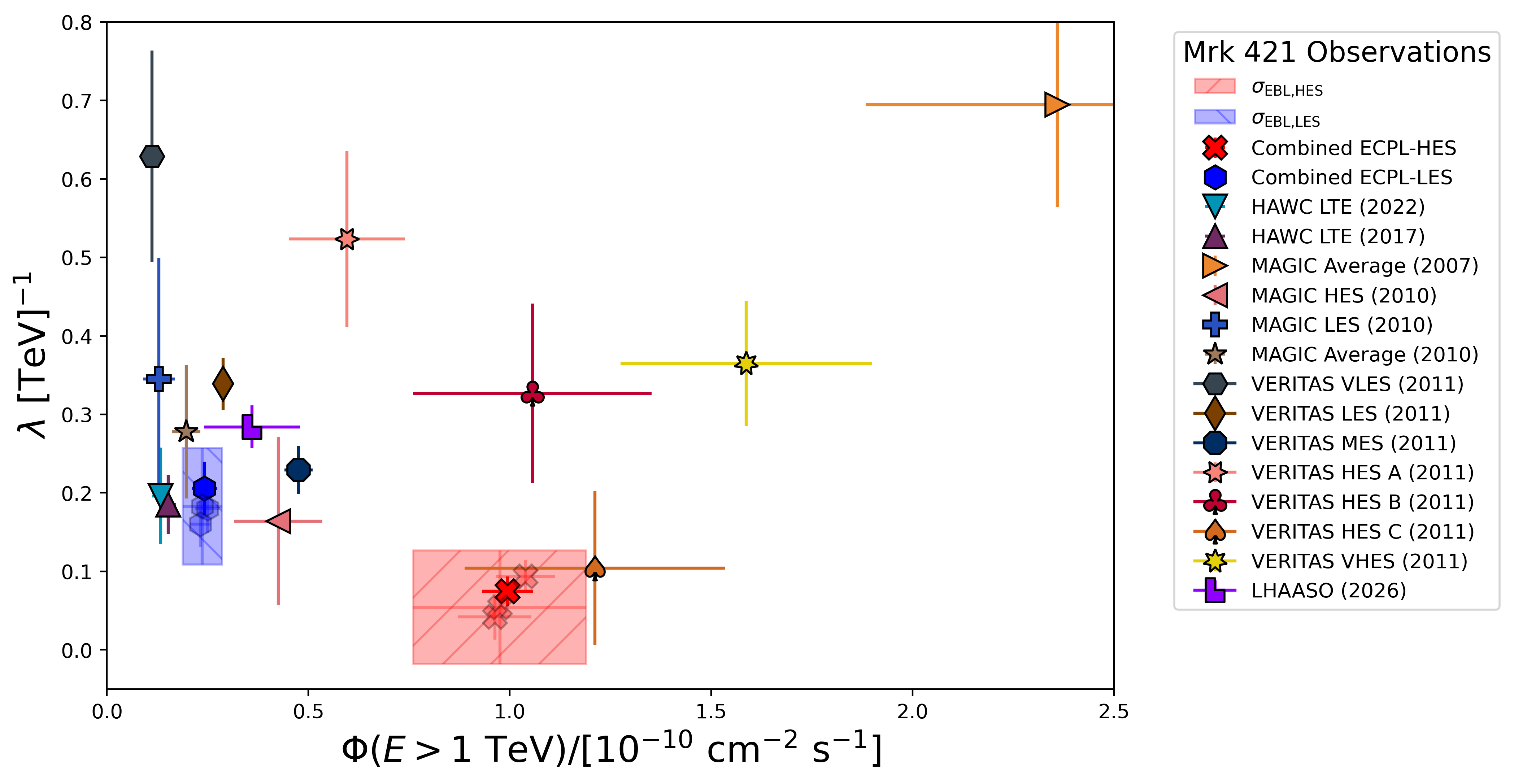}
    \caption{Comparison with results from previous observation campaigns. Upper panel. Scatter of emission states of Mrk 421 in the $(\alpha$--$\Phi)$ parameter space. Lower panel. Scatter of emission states in the $(\lambda$--$\Phi)$ parameter space. Data are taken from MAGIC \citep{magicMrk4212007,magicMrk4212010}, VERITAS \citep{veritasMrk4212011}, LHAASO \citep{lhaasomrk4212026}, and HAWC \citep{hawcMrk4212017,hawcMrk4212022} together with the results from this work.}
    \label{fig:comparison}
\end{figure}

\subsection{Spectral Hardening}

Several observation campaigns have been dedicated to observing Mrk 421 at different emission states, revealing a rather complex behavior in terms of spectral parameters and intensity of the emission. Following \citep{hawcMrk4212022} and our results of the best-fit model, the panels of Figure \ref{fig:comparison} show the comparison of our results with previous observation campaigns by MAGIC \citep{magicMrk4212007,magicMrk4212010} and VERITAS \citep{veritasMrk4212011}, and partial analysis by HAWC to estimate the long-term emission of Mrk 421 \citep{hawcMrk4212017,hawcMrk4212022}. We standardize the data presented in Figure \ref{fig:comparison} to be the observed flux in all cases. For studies that reported only the intrinsic spectrum, we recomputed the observed flux by convolving with the EBL model used in that study. While different models were used to estimate the spectral parameters in each study, this should have a small impact on the compilation plots, and trends (if any) should be preserved. The upper panel of Figure \ref{fig:comparison} shows the position of different observed emission states of Mrk 421 in the ($\alpha$--$\Phi$) parameter space. Qualitatively, and following the results from \citep{mrk421_20152016_low_emission}, the data indicates a hardening of the spectrum of Mrk 421 as a function of increasing $\Phi$. However, as $\Phi$ reaches values $\thicksim10^{-10}~\text{cm}^{-2}~\text{s}^{-1}$, an apparent flattening occurs, but a conclusive statement cannot be made because the data exhibits large scattering. As we mentioned in Section \ref{sec:results}, due to the limited size of our sample, we are not able to provide more evidence of the flattening. However, if such flattening is confirmed by upcoming analysis and observation campaigns, this would indicate that the processes responsible for injection no longer depend on the energy of the particles responsible for the observed gamma-ray emission \citep{mrk421_20152016_low_emission}.  On the other hand, the lower panel of Figure \ref{fig:comparison} shows the values of cutoff found across different observation campaigns. For simplicity, we only show the results of the joint spectra. Again, because of the large scattering in the data, it is hard to claim the existence of a possible evolution of $\lambda$ as a function of $\Phi$ in the compilation data.

We conclude by highlighting that long-term monitoring achieved by wide-field-of-view gamma-ray observatories provides crucial insights into the temporal behavior of extragalactic sources. It also corroborates the evidence of curvature or a cutoff in the spectrum of Mrk 421 and possible flattening of the spectral hardening. More analyses dedicated to Mrk 421 using HAWC data are in preparation.

\begin{acknowledgments}
    We acknowledge the support from: the US National Science Foundation (NSF); the US Department of Energy Office of High-Energy Physics; the Laboratory Directed Research and Development (LDRD) program of Los Alamos National Laboratory; Consejo Nacional de Ciencia y Tecnolog\'{i}a (CONACyT), M\'{e}xico, grants LNC-2023-117, 271051, 232656, 260378, 179588, 254964, 258865, 243290, 132197, A1-S-46288, A1-S-22784, CF-2023-I-645, CBF2023-2024-1630, c\'{a}tedras 873, 1563, 341, 323, Red HAWC, M\'{e}xico; DGAPA-UNAM grants IG101323, IG100726, IN111716-3, IN111419, IA102019, IN106521, IN114924, IN110521 , IN102223; VIEP-BUAP; PIFI 2012, 2013, PROFOCIE 2014, 2015; the University of Wisconsin Alumni Research Foundation; the Institute of Geophysics, Planetary Physics, and Signatures at Los Alamos National Laboratory; Polish Science Centre grant, 2024/53/B/ST9/02671; Coordinaci\'{o}n de la Investigaci\'{o}n Cient\'{i}fica de la Universidad Michoacana; Royal Society - Newton Advanced Fellowship 180385; Gobierno de Espa\~{n}a and European Union-NextGenerationEU, grant CNS2023- 144099; The Program Management Unit for Human Resources \& Institutional Development, Research and Innovation, NXPO (grant number B16F630069); Coordinaci\'{o}n General Acad\'{e}mica e Innovaci\'{o}n (CGAI-UdeG), PRODEP-SEP UDG-CA-499; Institute of Cosmic Ray Research (ICRR), University of Tokyo. H.M. acknowledges support under grant number CBF2023-2024-1630. H.F. acknowledges support by NASA under award number 80GSFC21M0002. C.R. acknowledges support from National Research Foundation of Korea (RS-2023-00280210). We also acknowledge the significant contributions over many years of Stefan Westerhoff, Gaurang Yodh and Arnulfo Zepeda Dom\'inguez, all deceased members of the HAWC collaboration. Thanks to Scott Delay, Luciano D\'{i}az and Eduardo Murrieta for technical support.
    This work was supported by the Natural Science Foundation of China with grant No. 12393853, the K. C. Wong Educational Foundation
\end{acknowledgments}

\section*{CRediT authorship contribution}
All authors are listed alphabetically according to the HAWC collaboration agreement. Authors S. Hern\`{a}ndez-Cadena, R. Torres-Escobedo, and H. Zhou led the analysis and writing.

\section*{Data Availability}

The data that support the findings of this article are available under request.

\bibliographystyle{aasjournalv7}
\bibliography{apssamp}

\end{document}